\documentclass[twocolumn]{aastex62}
\usepackage{natbib,tikz,amssymb,graphicx}
\setcitestyle{notesep={ }}
\tikzset{>=latex}
\usepackage{hyperref,verbatim}
\usepackage{amsmath}
\newcommand{\gp}{\ensuremath{\gamma_+}}
\newcommand{\gx}{\ensuremath{\gamma_\times}}

\begin{document}

\title{Precision Weak Gravitational Lensing Using Velocity Fields:
  Fisher Matrix Analysis}

\shorttitle{Precision Weak Lensing Using Velocity Fields}

\author[0000-0002-0813-5888]{David Wittman}
\affiliation{Physics Department, University of California, Davis, CA 
  95616}
\author{Matthew Self}
\affiliation{Physics Department, University of California, Davis, CA 
  95616}

\keywords{gravitational lensing: weak}

\begin{abstract} Weak gravitational lensing measurements based on
  photometry are limited by shape noise, the variance in the unknown
  unlensed orientations of the source galaxies.  If the source is a
  disk galaxy with a well-ordered velocity field, however, velocity
  field data can support simultaneous inference of the shear,
  inclination, and position angle, virtually eliminating shape
  noise. We use the Fisher Information Matrix formalism to forecast
  the precision of this method in the idealized case of a perfectly
  ordered velocity field defined on an infinitesimally thin disk.  For
  nearly face-on targets one shear component, $\gamma_\times$, can be
  constrained to $0.003\frac{90}{I_0}\frac{25}{n_{\rm pix}}$ where
  $I_0$ is the S/N of the central intensity pixel and $n_{\rm pix}$ is
  the number of pixels across a diameter enclosing 80\% of the
  light. This precision degrades with inclination angle, by a
    factor of three by $i{=}50^\circ$.  Uncertainty on the other shear
    component, $\gamma_+$, is about 1.5 (7) times larger than the
    $\gamma_\times$ uncertainty for targets at $i=10^\circ$
    ($50^\circ$).  For arbitrary galaxy position angle on the sky,
  these forecasts apply not to $\gamma_+$ and $\gamma_\times$ as
  defined on the sky, but to two eigenvectors in
  $(\gamma_+, \gamma_\times,\mu)$ space where $\mu$ is the
  magnification.  We also forecast the potential of less expensive
  partial observations of the velocity field such as slit
  spectroscopy.  We conclude by outlining some ways in which real
  galaxies depart from our idealized model and thus create random or
  systematic uncertainties not captured here. In particular, our
    forecast $\gamma_\times$ precision is currently limited only by
    the data quality rather than scatter in galaxy properties because
    the relevant type of scatter has yet to be measured.
\end{abstract}

\section{Introduction}\label{sec-intro}

Weak gravitational lensing is a key technique in modern cosmology, in
which the gravitational field of a celestial object is reconstructed
from the distortion it imprints on background sources of light; see
\citet{Bartelmann17} for a recent review.  The distortion is described
in terms of shear, defined as stretching the image in one direction
and compressing it in the perpendicular direction, and convergence,
defined as an isotropic stretching. Shear can be depicted as a
headless vector with a dimensionless magnitude and a position angle
(PA) on the sky modulo 180$^\circ$, or in terms of two components
separated by 45$^\circ$ in PA.  Shear is inferred from the observed
shapes of source galaxies, under the assumption that galaxies have no
preferred orientation in the absence of lensing. The fundamental
source of noise in this approach is the large intrinsic {\it scatter}
in galaxy orientations, called shape noise. This scatter is such that
the shear on a single galaxy is uncertain by at least $0.2$ in each
component, while the relevant signal is usually much
smaller. Averaging over many source galaxies in a given patch of sky
builds the signal-to-noise ratio (S/N), but correspondingly decreases
the angular resolution of the reconstruction.

Techniques to measure convergence also face substantial amounts of
noise. Convergence leads to magnification, which increases the flux of
sources while decreasing the effective area of sky probed. This can
shift the counts of sources as a function of apparent magnitude
\citep[eg,][]{Morrison12,DESSVmag}. This is again a technique that
relies on aggregation of many sources due to the low information
content of each individual source.

To increase the information content of an individual source, we must
know more about its unlensed state. A recent idea in this regard is
that a source with a well-ordered velocity field, such as a rotating
disk galaxy, can potentially provide that information. The velocity in
each pixel provides a tag that helps place that pixel in the source
plane---a more specific tag than is possible with the intensity
field. Although velocity measurements are more expensive than
intensity measurements, the gain in per-galaxy precision is
potentially quite large.  This paper aims to quantify that gain with a
Fisher information matrix analysis.

First, we briefly outline the history of the velocity field idea.
\citet{Blain2002} first recognized that shear perturbs the symmetry of
the velocity field. He used a rotating ring toy model to show how
velocity measurements could constrain the shear component at
$45^\circ$ to the source galaxy's unlensed photometric axes, which
we call \gx. \citet{Morales2006} extended the velocity-field idea to
full disk galaxies, and provided a clear picture of how \gx\ causes
the major and minor velocity axes to deviate from perpendicularity.  A
version of this method has been implemented by \cite{deBurghDay2015},
who infer the shear by determining the transformation required to
restore symmetry to the velocity map. They find that shears as small
as 0.01 are measurable in simulations, and they find shears consistent
with zero, with uncertainties $\sim 0.01$, on unlensed nearby disk
galaxies.  However, their approach is still insensitive to the
component of shear along the unlensed photometric axes because that
component, which we call \gp, preserves the symmetry of the velocity
field.

\gp\ does change the observed axis ratio, so \citet{Huff2013}
proposed constraining this component as follows. They propose
predicting the total rotation speed of the galaxy using the Tully-Fisher
relation \citep{TullyFisher}, then comparing this prediction with the
measured line-of-sight rotation speed to find the inclination of the
disk. Assuming the disk to be circular when viewed face-on, the
inclination uniquely predicts the unlensed axis ratio, which
effectively removes the problem of shape noise.  The \citet{Huff2013}
goal of designing an efficient large cosmic shear survey led them to
propose minimal velocity-field measurements per galaxy (slit spectra
along the apparent photometric axes) and to assume approximations,
such as the low-shear limit and negligible magnification, that may
fail in more general lensing situations. Considering that
\cite{deBurghDay2015} needed the full velocity field of a very
well-resolved nearby galaxy to infer \gx, it is not clear
that shear could be measured precisely using only crossed slits along
the photometric axes. Nevertheless, the insight of
\citet{Huff2013}---that symmetry is not the only source of information
in the velocity field---is potentially powerful and deserves further
investigation.

This paper uses the Fisher Information Matrix formalism to forecast
the best achievable performance in the case of perfectly ordered
rotation and an infinitesimally thin disk.  This is highly idealized,
but the point is to determine whether the method is promising enough
to justify further development.  Therefore, we forecast the best
possible performance across a wide range of scenarios: from zero-shear
lines of sight on up to higher-shear lines of sight, from nearly
face-on targets to nearly edge-on targets, from full velocity-field
observations to crossed slits and so on.

The remainder of this paper is organized as follows.  In
\S\ref{sec-method} we describe and illustrate the method; in
\S\ref{sec-results} we present the resulting forecasts; and in
\S\ref{sec-discussion} we discuss the implications.

\section{Method}\label{sec-method}

We assume an infinitesimally thin disk galaxy with a polar $(R,\phi)$
coordinate system specifying particle locations.  Viewed at
inclination $i$ (where $i=0$ is face-on) but before lensing, we define
an $(x,y)$ coordinate system, in which
\begin{eqnarray}
x&=& R\cos(\phi-\phi_0) \cos i\\
y&=& R\sin(\phi-\phi_0)
\end{eqnarray}
where $\phi_0$ is the unlensed PA of the apparent major
axis.  The velocity field is assumed to be a function only of $R$,
with measured line-of-sight velocity
$v_{los}=v(R)\sin(\phi-\phi_0)\sin i$.

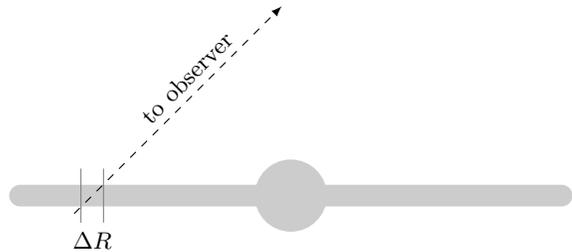
\begin{figure}
\centerline{\begin{tikzpicture}[scale=1.2]
\fill[black!20] (-3,-0.12) -- ++(6,0) arc(-90:90:0.12) -- ++(-6,0) arc(90:270:0.12);
\coordinate (P) at (-2.2,0);
\draw[dashed,->] (P) ++(-0.2,-0.2) to node[pos=0.6,sloped,above] {to
  observer} +(2.3,2.3);
\draw[gray] (P) ++(-0.125,-0.3) -- +(0,0.6);
\draw[gray] (P) ++(0.125,-0.3) -- +(0,0.6);
\node at (-2.2,-0.5) {\small $\Delta R$};
\fill[black!20] (0,0) circle(0.4);
\end{tikzpicture}}
\caption{Effect of finite disk thickness. A line of sight probes
  particles at a range of cylindrical galactocentric distances $R$
  depending on their height above or below the midplane. Where the
  rotation curve is approximately linear in $R$ across the range
  $\Delta R$, the above- and below-plane contributions are
  approximately equal and opposite, which preserves the mean velocity
  but increases the linewidth.  Hence to first order the disk can be
  modeled as an infinitesimally thin disk but with greater
  linewidth.}\label{fig-thick}
\end{figure}

Note that, to first order, a finite-thickness disk can be modeled as
an infinitesimally thin disk but with greater
linewidth. Figure~\ref{fig-thick} illustrates the argument: stars
along the line of sight above and below the disk depart from the
midplane value of $R$ in equal and opposite ways.  Therefore the mean
velocity for this line of sight is unchanged if the rotation curve is
linear in $R$ across the range of $R$ probed by the line of sight. The
line of sight does, however, encounter a wider range of velocities
than would be the case for an infinitesimally thin disk, leading to a
greater linewidth unless the rotation curve is approximately flat
across the range of $R$ probed by a given line of sight.  Real
galaxies will present additional complications, such as bulges and
warps. We stress that our approach here is to explore the optimal case
of a bulgeless, dynamically cold thin disk in order to establish the
limits of this method, reveal parameter degeneracies and requirements
for priors, and identify key assumptions that will need to be explored
further.

Lensing transforms the coordinates described above to observed
coordinates, which we denote with primes:
\begin{align}
\begin{bmatrix} x^\prime \\ y^\prime \end{bmatrix} = A^{-1} \begin{bmatrix} x \\ y \end{bmatrix}
\end{align}
where
\begin{align}
A^{-1} = \mu \left( \begin{matrix} 1-\kappa+\gamma_+ & -\gamma_\times \\ -\gamma_\times & 1-\kappa-\gamma_+ \end{matrix} \right)\label{eqn-A}
\end{align}
Here $\kappa$ is the convergence, which is proportional to the surface
mass density; $\mu = \frac{1}{(1-\kappa)^2-\gamma^2}$ is the
magnification, and $\gamma = \sqrt{\gamma_+^2+\gamma_\times^2}$ is the
magnitude of the shear. We choose to parametrize the shear in terms of
\gp\ and \gx, which are dimensionless quantities with identical
ranges, rather than a magnitude and a PA. Then, the lensing matrix can
be completed by specifying either $\kappa$ or $\mu$. We choose $\mu$
because prior information on $\mu$ is more likely to be available
through other methods.

\begin{figure}
\centerline{\includegraphics[scale=0.65]{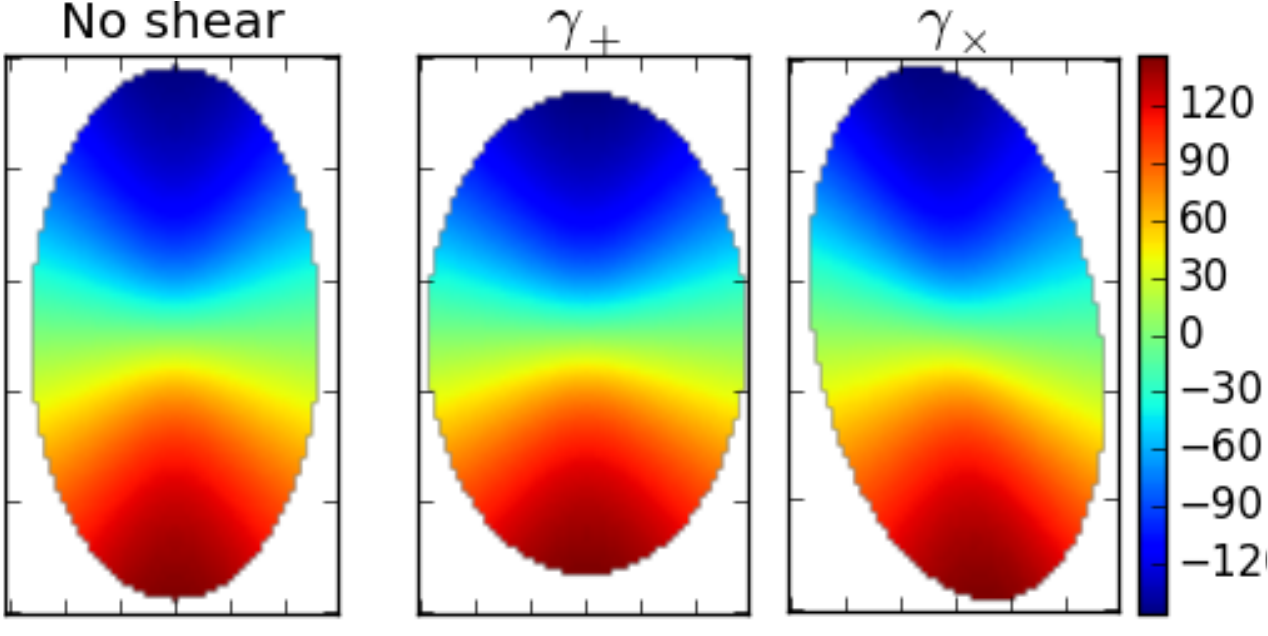}}
\caption{Velocity fields before lensing (left), after
  applying $\gamma_+=0.1$ (middle), and after
  applying $\gamma_\times=0.1$ (right). The galaxy has maximum
  rotation speed of 220 km/s and is inclined at 1 radian to the line
  of sight. The colorbar shows units of km/s.}\label{fig-vfieldcolor}
\end{figure}

With this in mind, the left panel of Figure~\ref{fig-vfieldcolor}
shows an unlensed model velocity field for $i=60^\circ$ and
fortuitously aligned with the coordinate axes.  The middle and right
panels show the same field after lensing by $\gamma_+$ and
$\gamma_\times$ respectively. (All fields in this figure are cropped
at a consistent physical radius; this guides the eye but may overstate
the power of the method, because such cuts and comparisons will not be
available to the data analyst.)  The right panel displays the
asymmetry discussed in the introduction, which we will associate with
$\gamma_\times$ throughout the paper.  Our formalism defines the shear
components with respect to sky coordinate axes rather than the galaxy
axes, so in practice the asymmetry-causing component need not be
$\gamma_\times$ as defined on the sky. Although the physical
distinction is between shear components aligned and not aligned with
the apparent unlensed galaxy axes, we choose not to {\it define}
the components this way because in practice the unlensed axes are
unknown. By defining shear components on the sky, we adopt the basis
in which shear will actually be used.  That said, to highlight
physical behaviors we will typically align the galaxy as in
Figure~\ref{fig-vfieldcolor} and refer to $\gamma_\times$ as causing
the asymmetry.

\begin{figure*}
\hskip1.4in{\large Velocity}\hskip3in{\large Intensity}\\
\rotatebox{90}{\hskip1.1in\large $i=20^\circ$}\hskip-6pt
\centerline{\framebox{\includegraphics[width=3.3in]{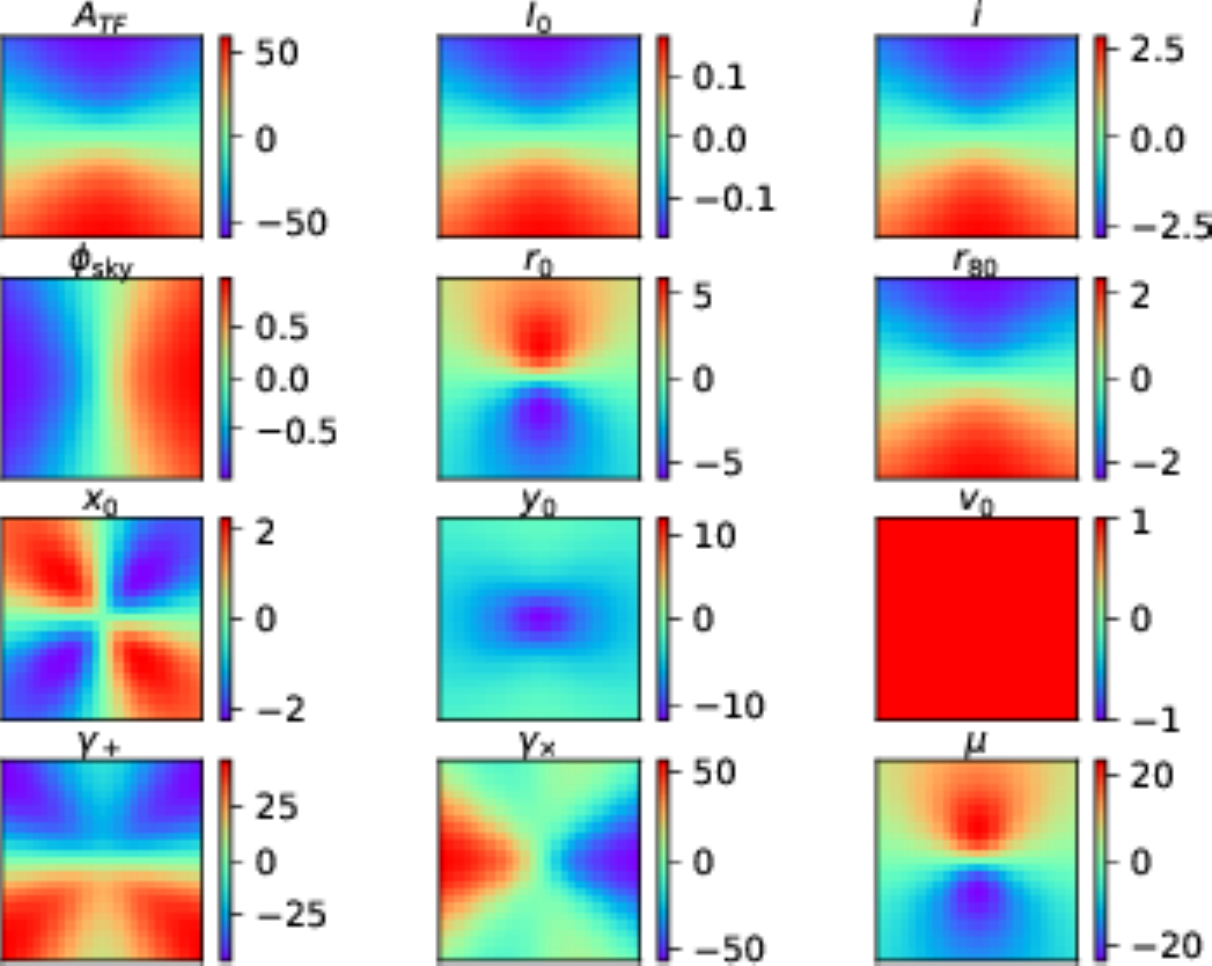}}
\framebox{\includegraphics[width=3.3in]{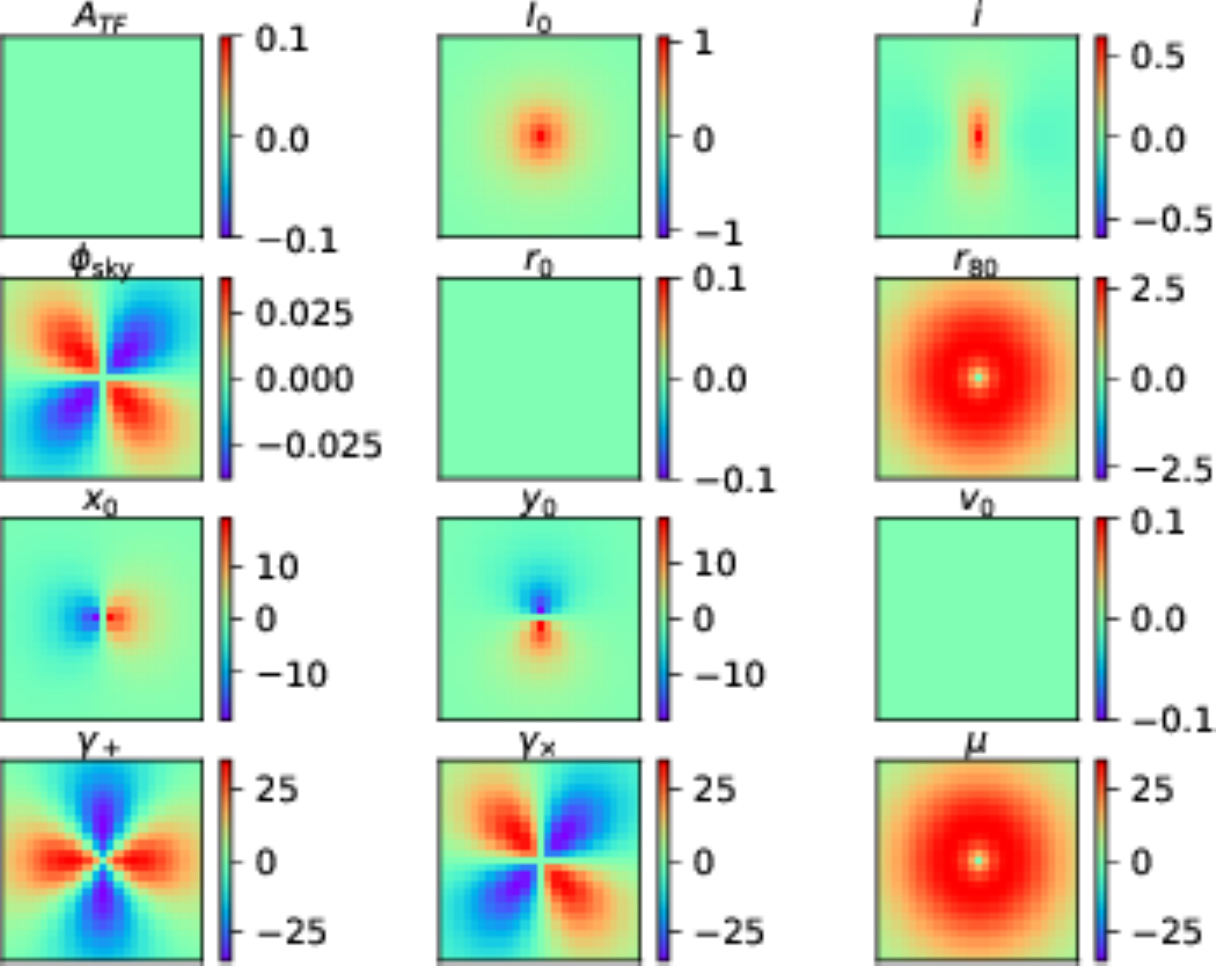}}}
\rotatebox{90}{\hskip1.1in\large $i=60^\circ$}\hskip-6pt
\centerline{\framebox{\includegraphics[width=3.3in]{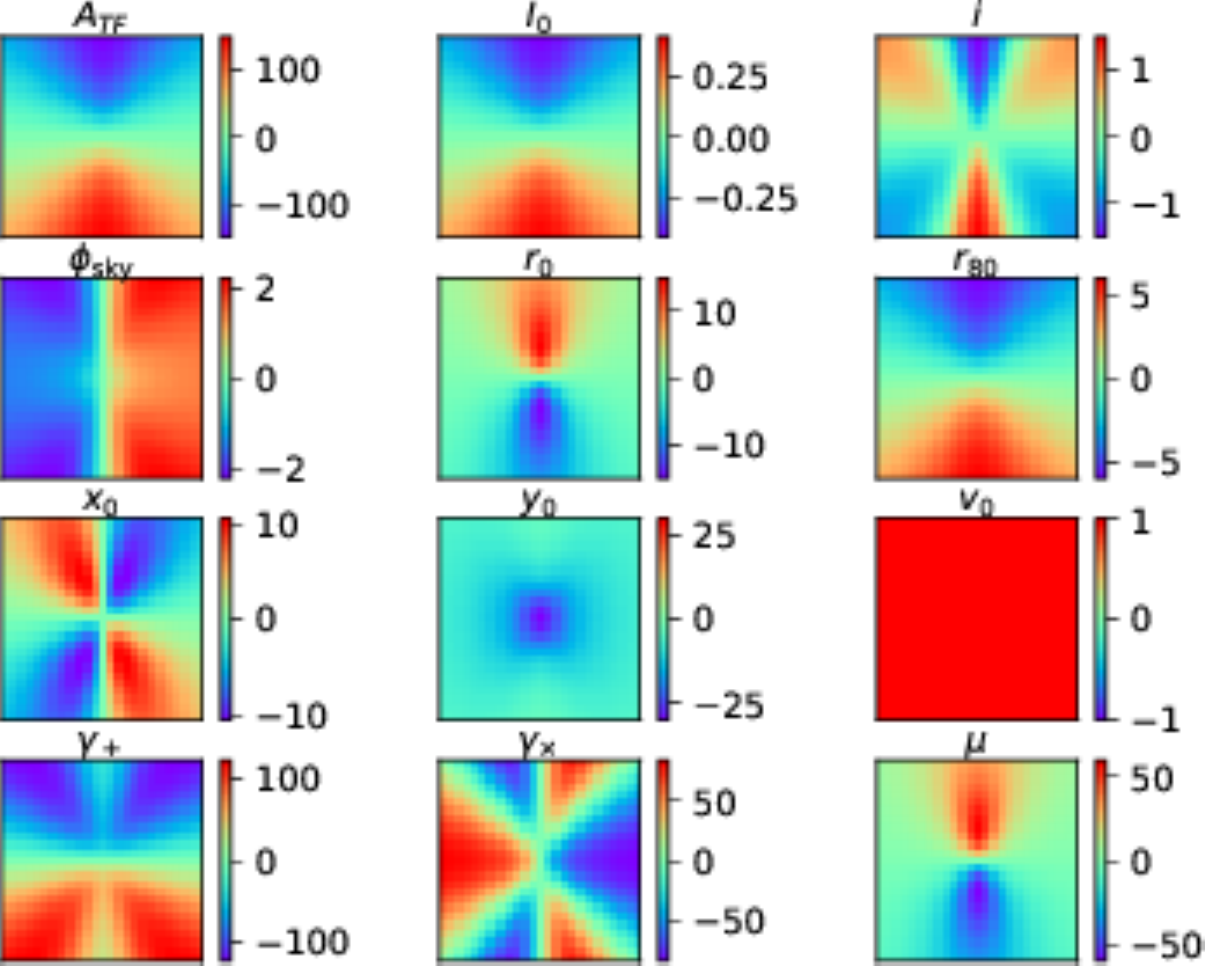}}
\framebox{\includegraphics[width=3.3in]{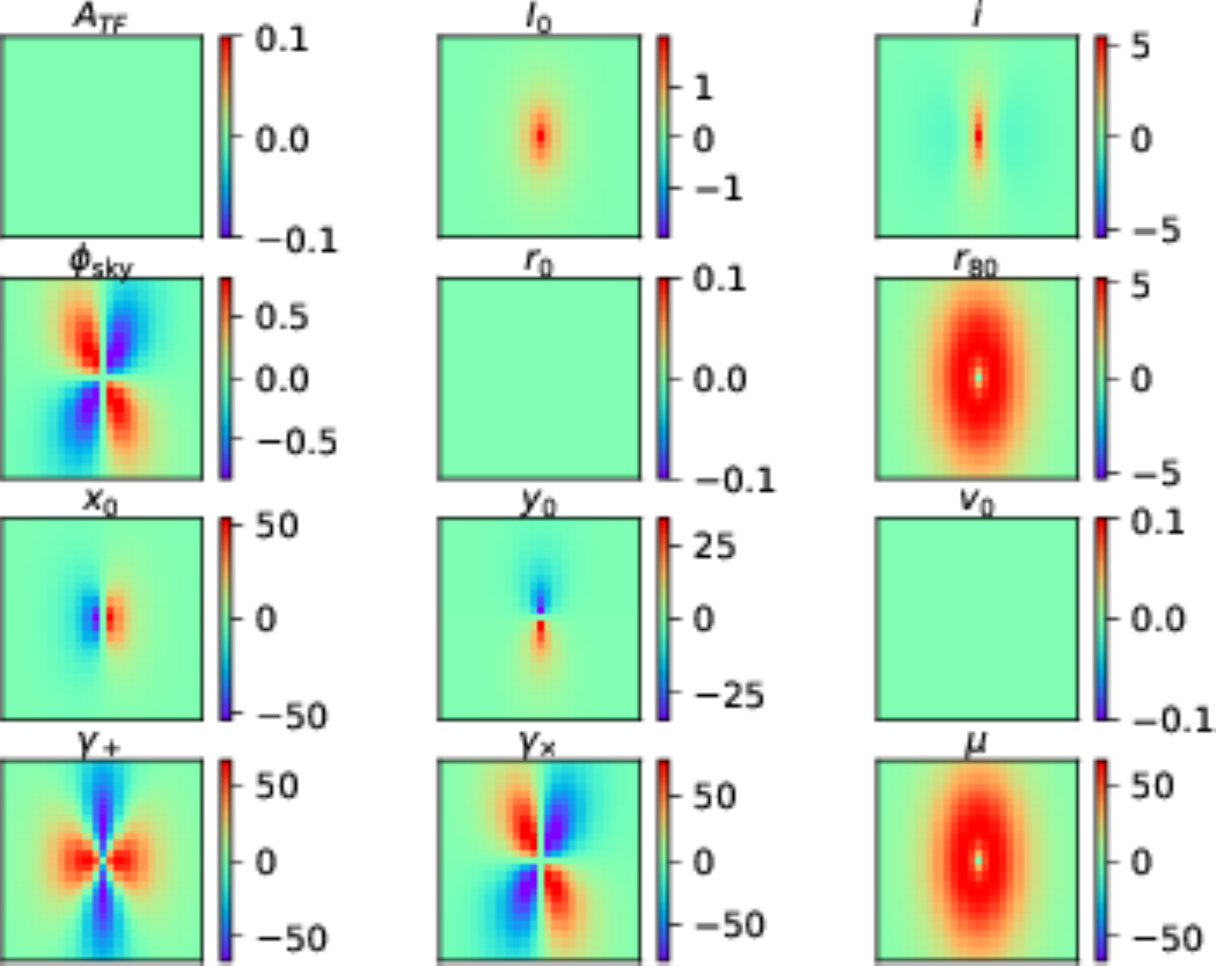}}}
\caption{Partial derivatives of the velocity (left) and intensity
  (right) fields with respect to each of the parameters, for an
  inclination of 20$^\circ$ (top row) and 60$^\circ$ (bottom row).
  The colorbar units are km/s on the left, and arbitrary intensity
  units on the right. To keep the scales roughly the same across
  panels, we show the change in velocity per 0.01 change in shear and
  convergence. The form of the velocity field itself can be seen in
  the $A_{TF}$ panels on the left because that field is linear in $A_{TF}$.
  Similarly, the form of the intensity field can be seen in
  the $I_0$ panels on the right.}\label{fig-partials}
\end{figure*}  

A key assumption is that the unlensed velocity field has the
symmetry shown. Under this assumption, the data analyst can determine
$\gamma_\times$ because the relevant unlensed condition is known.
The effect of $\gamma_+$ is to change the apparent axis ratio, so
measuring $\gamma_+$ requires knowledge of the unlensed axis
ratio. That axis ratio is set by the inclination, an effect distinct
from that of $\gamma_+$ in that inclination {\it also} changes the
line-of-sight velocity. It is conceptually useful to consider the
extreme case of a uniform observed velocity field, from which we can
deduce that the galaxy must be viewed face-on. This implies a
unlensed axis ratio of unity, so we can deduce $\gamma_+$ from the
observed axis ratio, with no shape noise.\footnote{In practice, there
  will still be some uncertainty due to uncertainty in the intrinsic
  circularity of face-on disks.}  The key is the ability to deduce a
unlensed axis ratio from the velocity field amplitude; this is a
way of restating the idea of \citet{Huff2013}.

To go beyond this conceptual understanding we must choose quantitative
models for the intensity and velocity fields.  First, we define the
parameter $r_{80}$, which is the radius that encircles 80\% of the
galaxy light. For an exponential disk, this is 2.99 times the
exponential scale length. The intensity field is specified by
$I=I_0\exp(-\frac{2.99R}{r_{80}})$, where the parameter $I_0$
represents the central intensity. We set the intensity uncertainty in
each pixel to unity, so $I_0$ represents the S/N of the intensity
measurement in the central pixel.  The intensity uncertainty field is
uniform because sky noise, rather than photon noise from the galaxy
itself, is the dominant uncertainty in broadband imaging of most
galaxies. We set the fiducial value of $I_0$ to 90, which is a high
S/N reflecting the fact that bright galaxies are the likeliest targets
for integral field spectroscopy.  The velocity uncertainty is set by
$\sigma_{v,0}$, the uncertainty in the central pixel (with a fiducial
value of 10 km/s) and grows exponentially with $R$ because source
photon noise is likely to be the limiting factor.

We adopt a simple arctan rotation curve:
$v=v_{\rm max} \frac{2}{\pi} \arctan \frac{R}{r_{0}}$, where the
factor $\frac{2}{\pi}$ ensures that $v\to v_{\rm max}$ as $r\to\infty$
given an arctan function that returns radians.  We also investigated
the more complicated Universal Rotation Curve
\citep[URC;][]{Persic96,Salucci2007} and found the results to be
nearly identical; a few minor differences will be discussed in \S\ref{sec-URC}.
With either form, the rotation curve has a scale length independent of
the scale length describing the intensity field. If these two scales
were the same, the model would be more constrained and yield higher
precision, but the scales do appear to differ in observed galaxies.

$v_{\rm max}$ is related to the intensity field via the Tully-Fisher
relation (TFR) as follows. The TFR empirically states that
$L\propto v_{\rm max}^n$ where $n\approx 4$, with a scatter in
luminosity or stellar mass of about 16\% \citep{MillerTFR2011}. This implies that
at fixed $L$ the scatter in $v_{\rm max}$ is about 4\%. For an
exponential disk, the total luminosity is $L\propto I_0r_{80}^2$, so
the TFR predicts $v_{\rm max} \propto (I_0r_{80}^2)^{0.25}$.  With our
fiducial values of $I_0$ and $r_{80}$ (12.5 pixels), we need
$v_{\rm max} =20 (I_0r_{80}^2)^{0.25}$ to produce a typical rotation
speed around 200 km/s.\footnote{More precisely, $v_{\rm max}=218$ in this case,
but note that in the arctan model $v_{\rm max}$ is reached only as
$R\to\infty$.} Hence we define a Tully-Fisher amplitude, $A_{TF}$, with a
fiducial value of unity, such that
$v_{\rm max} = 20 A_{TF} (I_0r_{80}^2)^{0.25}$.  We then place a
prior of $\pm 4\%$ on $A_{TF}$.

Table~\ref{tab-params} summarizes the parameters for this model,
including the nuisance parameters $x_0,y_0,v_0$ describing the galaxy
position on the sky and systemic radial velocity.  The units listed in
this table are relevant to the forecast precision plots presented
below; units are omitted for dimensionless quantities.  The results
can be quite sensitive to the inclination angle $i$, so $i$ will be
varied in many plots rather than remaining fixed at a fiducial value.

\begin{table*}
\centering
\caption{Model parameters}
\begin{tabular}{cccc}
\hline
\hline
\multicolumn{1}{c}{Symbol} & \multicolumn{1}{c}{Fiducial value}&
\multicolumn{1}{c}{Unit}  &\multicolumn{1}{c}{Description} \\ \hline
$A_{TF}$ & 1 &-& $v_{\rm max}$ as a fraction of the Tully-Fisher prediction\\
$I_0$ & 90 &- & intensity S/N at center\\
$i$	& varies & deg & inclination angle \\
$\phi_{\rm sky}$ &  0 & deg& sky position angle of unlensed major axis\\
$r_{0}$	& 4 & pixel & rotation curve scale length\\
$r_{80}$	& 12.5 &pixel & radius of 80\% encircled light\\
$x_{0}$	& 0 & pixel & center of galaxy in $x$ coordinate\\
$y_{0}$	& 0 & pixel & center of galaxy in $y$ coordinate\\
$v_{0}$	& 0 & km/s & galaxy systemic radial velocity\\
$\gamma_+$	& 0 &- &shear parallel to sky coordinates \\
$\gamma_\times$ & 0 &-& shear at 45$^\circ$ to sky coordinates\\
$\mu$	& 1 &-& magnification \\
\hline\hline
\multicolumn{4}{c}{Data parameters}\\
\hline
$n_{\rm pix}$ & 25 & pixel & field diameter \\
$\sigma_{v,0}$ & 10 & km/s & uncertainty in $v$, central pixel \\
\end{tabular}
\label{tab-params}
\end{table*}

We construct velocity and intensity fields extending to a radius of
$r_{80}$, thus encompassing 25$\times25$ pixels each.  We compute
partial derivatives numerically to a relative precision of order
$10^{-11}$  using the algorithm in Section 5.7 of \citet{NumRec},
which we re-implement in Python.  Figure~\ref{fig-partials} shows the
partial derivatives of the velocity and intensity fields with respect
to each parameter at two different inclinations. These figures will
help readers understand which parameters are highly correlated. Note
that $A_{TF}$, $I_0$, $i$, and $r_{80}$ have nearly identical effects
on the velocity field. For $A_{TF}$ this is broken by its lack of
effect on the intensity field, but $I_0$ and $r_{80}$ also have nearly
identical effects on that field---with opposite sign, but the sign is
not relevant for determining degeneracy and correlation. The effect of
$i$ on the intensity field is not identical to that of $I_0$, but
there is a good deal of overlap, indicating that the three parameters
$I_0$, $i$, and $r_{80}$ will be highly correlated.  Magnification
($\mu$) joins this family because its effect on both velocity and
intensity fields is much like $-I_0$, and its effect on the intensity
field is identical to changing the intensity scale length $r_{80}$.
Finally, $r_0$ is linked with all these parameters because, as a
rotation curve scale length, its effect on the velocity field is
identical to that of magnification $\mu$.  The strength of these
correlations will vary with the specific values of inclination, shear,
and so on: Figure~\ref{fig-partials}, for example, shows that by
$i=60^\circ$ perturbations in $i$ affect the velocity field
differently than perturbations in $I_0$ and $r_{80}$.


For any given value of $i$, we concatenate the velocity and intensity
fields into a Python data structure representing a
generalized data field we denote $\vec{D}$. Denoting the set of
parameters as $P$, the Fisher matrix elements are then
\begin{equation}
\mathcal{F}_{ij} = \sum_{\rm pixels}\vec{\sigma}^{-2}(\frac{\partial \vec{D}}{\partial P_i})(\frac{\partial \vec{D}}{\partial P_j}) \label{eqn-FIM}
\end{equation}
where $i$ and $j$ index the parameters, and $\vec{\sigma}$ is the
uncertainty field associated with the data field.  We then invert the
Fisher matrix to obtain the covariance matrix $C$. We also compute the
correlation matrix $\rho\equiv D^{-1}CD^{-1}$ where
$D\equiv\sqrt{\textrm{diag}(C)}$.

\section{Results}\label{sec-results}

\subsection{Degeneracies}

We find that $\frac{\partial \vec{D}}{\partial \gamma_+}$ is a linear
  combination of the other partial derivative fields. The
  coefficients depend on the parameter values themselves, but for
  concreteness we display the coefficients for our fiducial scenario
  at $i=30^\circ$:
  \begin{equation}
    \label{eq-degen}
 \frac{\partial
   \vec{D}}{\partial \gamma_+} +199\frac{\partial \vec{D}}{\partial i}
 -180\frac{\partial \vec{D}}{\partial I_0}
+4\frac{\partial \vec{D}}{\partial r_0}
+12.5\frac{\partial \vec{D}}{\partial r_{80}}
 -6\frac{\partial \vec{D}}{\partial A_{\rm TF}}=0
  \end{equation}
    Hence, a
model can be transformed into another model with a different
$\gamma_+$ value that predicts the same data, providing that we:
\begin{itemize}
\item increment $i$ to preserve the apparent axis ratio despite the
  change caused by $\gamma_+$. (In our setup, the unlensed apparent
  major axis is in the ``$y$'' direction while positive $\gamma_+$
  acts to stretch the ``$x$'' direction, hence one must make the galaxy
  more edge-on to counteract positive $\gamma_+$.)
\item decrement $I_0$ to preserve the apparent surface brightness. (In
  our model the galaxy is transparent, so making it more edge-on
  had the side effect of increasing the apparent surface brightness.)
\item increment $r_{80}$ to preserve the observed angular size of the
  major axis of the intensity field. In concert with the change in
  inclination angle, this also preserves the apparent minor axis.
\item increment $r_{0}$ to preserve the observed angular scale of the
  rotation curve's rise.  The coefficients on $r_{80}$ and $r_{0}$
  here are equal to their fiducial values, confirming that $d\gamma_+$
  equals the {\it fractional} change in each apparent size, as it
  should when $\kappa=0$.
\item finally, we have to preserve the Tully-Fisher relation. To
  preserve the amplitude of the observed velocity-field pattern
  despite being more edge-on, our model must suppose a lower rotation
  speed, thus decreasing $A_{\rm TF}$. Alternately, the same effect
  can be achieved by adjusting $\mu$, which allows a lower-luminosity
  model to fit the intensity field.
\end{itemize}
The specific linear combination depends on the scenario, but always
involves $A_{\rm TF}, \mu, I_0, i, r_0,$ and $r_{80}$. It also
involves $\phi_{\rm sky}$ if the $\gamma_\times$ is nonzero, and
$\gamma_\times$ if $\phi_{\rm sky}$ is nonzero.  We tested a
parametrization in terms of reduced shear
($g_{+(\times)}\equiv\frac{\gamma_{+(\times)}}{1-\kappa}$) rather than
shear. This did not change the set of interdependencies.  We find
similar dependencies parametrizing in terms of $\kappa$ rather than
$\mu$.  This degeneracy prevents the Fisher matrix from being
inverted.

The TFR should be effective at breaking
this degeneracy, in the sense that a fractional change in $\gamma_+$
requires a large fractional change in $A_{\rm TF}$, in tension with
the TFR. But the fact that a tweak in $\mu$ can substitute for a tweak
in $A_{\rm TF}$ leaves the Fisher matrix still
noninvertible.

However, when {\it finite} steps are taken along the degeneracy
direction, the data change quadratically with the step size,
suggesting that data can indeed constrain the model. We have confirmed
this with Markov Chain Monte Carlo (MCMC) explorations of the
likelihood surface: our fiducial data constrain $\mu$ to slightly
better than $\pm0.1$ at all inclinations.  Hence we support the Fisher
forecast by placing a prior of $\pm0.1$ on $\mu$.  The fact that data
constrain $\mu$ as well as all the other parameters is potentially
important and will be further explored in a subsequent paper using
higher order expansions of the likelihood surface \citep{Heavens2016}
and/or MCMC techniques.

The physical context is that $\mu=1$ in the absence of lensing; only
the densest lines of sight have $\mu$ approaching 2 or more; and for those lines of
sight the fact that $\mu$ is high will generally be known in
advance.  We also note that the weak lensing formalism used here
breaks down at high magnification. Specifically, we assume that the
matrix $A$ (hence the parameters $\gamma_+$, $\gamma_\times$, and
$\mu$) is constant over the extent of the target galaxy, and this is
not generally the case along strongly lensed lines of sight. In those
cases, more traditional strong-lensing techniques will be preferred,
although it is possible that the velocity field can complement the
intensity field in constraining the strong-lensing reconstruction
\citep{Rizzo2018}.  For all these reasons, the $\mu$ uncertainty
  on a typical weak lensing
  line of sight would approach 0.1 in any case, so our prior is a good
  match to the physical situation.
Table~\ref{tab-priors} lists the priors applied as part of our
standard forecast.

\begin{table}
\centering
\caption{Priors}
\begin{tabular}{cc}
Parameter & Width (Gaussian $\sigma$) \\ \hline
$A_{TF}$ & 0.04 \\
$\mu$ & 0.1 \end{tabular}
\label{tab-priors}
\end{table}

With this prior in place, we invert the Fisher matrix.  Numerical
instability in matrix inversion generally becomes important if the
inverse of the condition number, the ratio between largest and
smallest modulus eigenvalue, is not much larger than the inaccuracies
in our knowledge of the matrix elements \citep{2008PhRvD..77d2001V}.
With the $\mu$ prior we find condition numbers ranging from
  $\approx10^{6.5}$ at $i=10^\circ$ to $\approx10^{11.5}$ at
  $i=80^\circ$.  The inverse of the latter overlaps the
  $\sim10^{-11}$ uncertainty in our numerical differentiation cited
  above.\footnote{For standard double precision arithmetic, the
    relative rounding error is $\approx 10^{-16}$, so machine
    precision is a subdominant uncertainty here.}  Hence, we cannot
  make numerically stable forecasts for galaxies close to edge-on. We
  limit our forecasts to those with condition number $<10^8$
  ($i\le 50^\circ$).  Appendix~\ref{appendix} shows that forecasts
  with these condition numbers match very well with MCMC
  explorations of the likelihood surface.  In the verified results
  below, the shear constraints consistently degrade as the inclination
  increases from $10^\circ$ to $50^\circ$, so there is little reason
  to push the forecast to higher inclination.


\begin{figure*}
 \centerline{\includegraphics[width=3.3in]{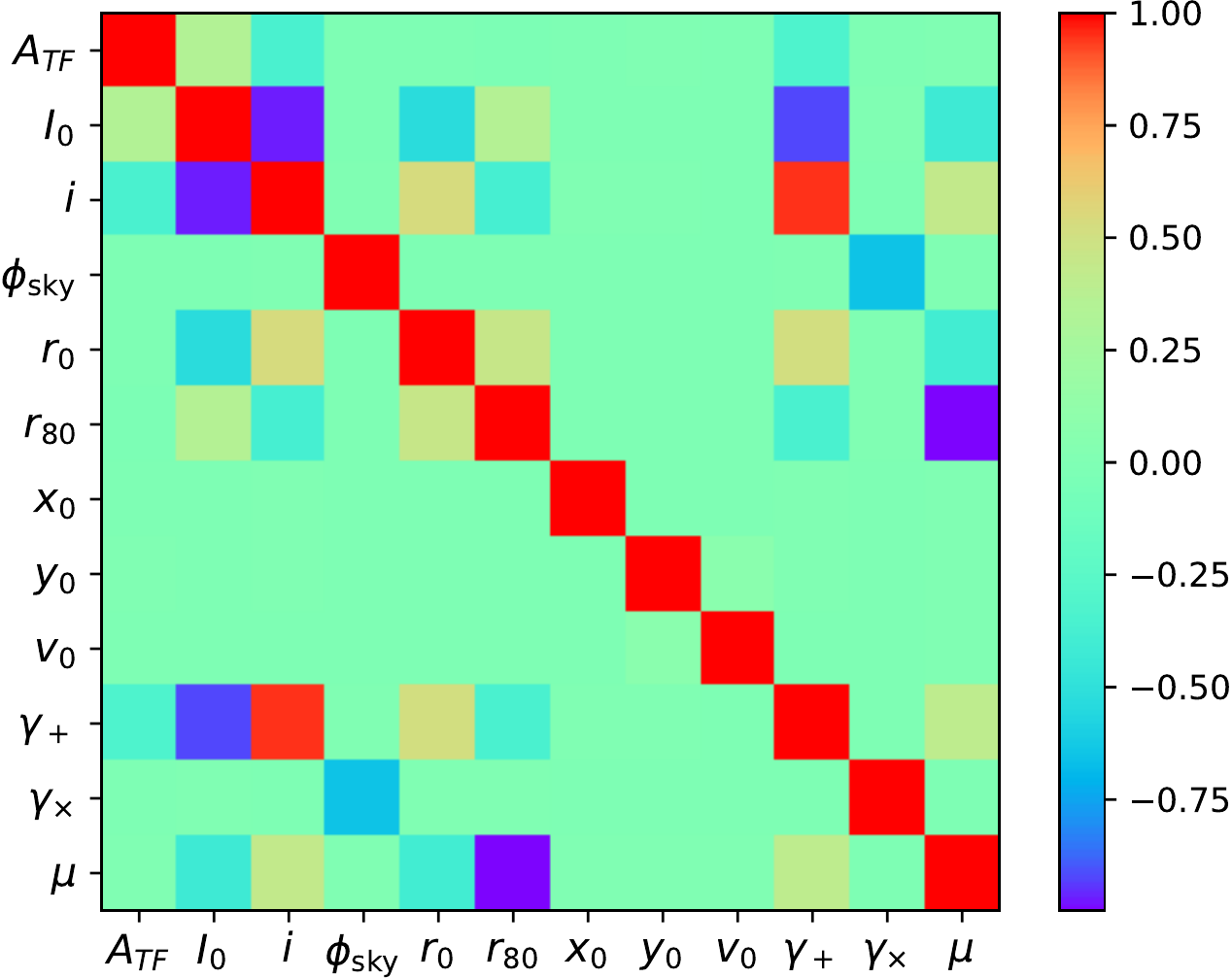}\hskip1cm\includegraphics[width=3.3in]{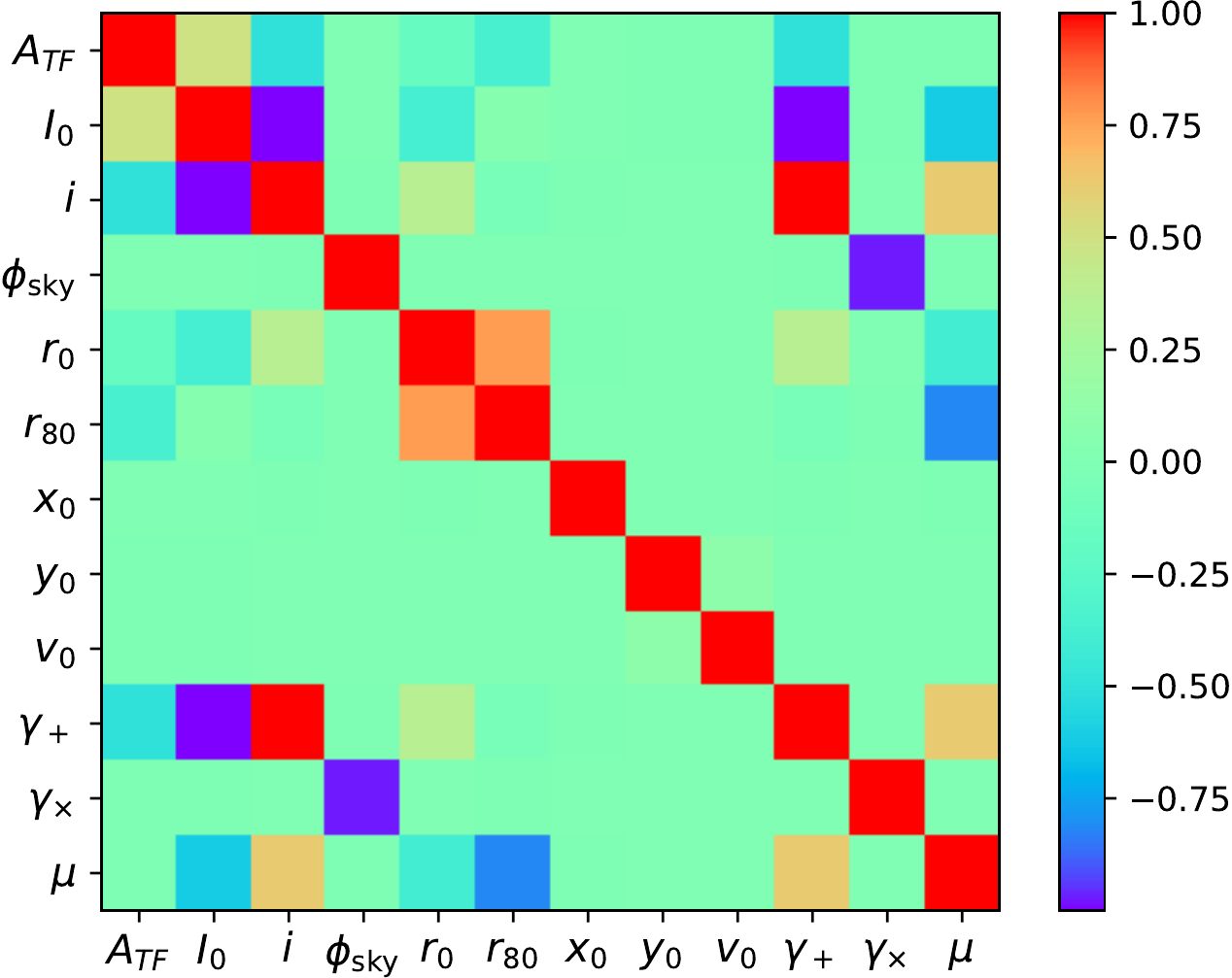}}
 \caption{Correlation matrices for the $i=20^\circ$ (left) and
   $i=50^\circ$ (right) cases. \label{fig-corr}}
\end{figure*}

Figure~\ref{fig-corr} shows the resulting correlation matrices for
low- and high-inclination cases.  The parameters in the family
  discussed above ($A_{\rm TF}$, $I_0$, $i$, $r_0$, $r_{80}$,
  $\gamma_+$, and $\mu$) are indeed correlated, with some increase in
  correlation at higher inclination.  Separately, there is an
  anticorrelation between $\phi_{\rm sky}$ and $\gamma_x$, which is
  moderately strong at $i=20^\circ$ but quite strong at
  $i=50^\circ$. This suggests that higher inclinations will yield
  looser constraints for both shear components. These correlations
set the stage for understanding our primary products, forecasts of
precision on each parameter.

\begin{figure}
 \centerline{\includegraphics[width=\columnwidth]{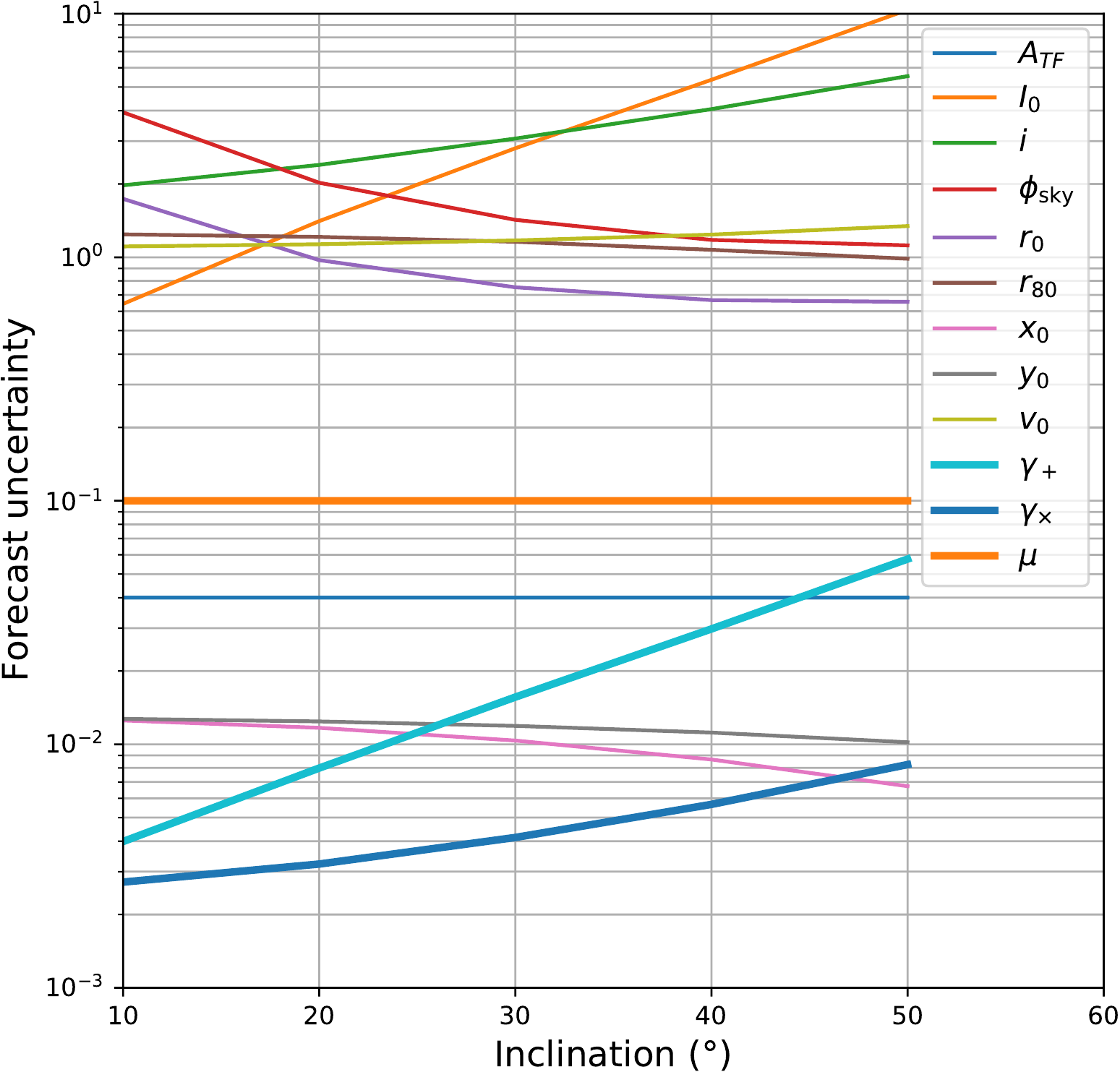}}
 \caption{Forecast constraints as a function of inclination angle $i$.}\label{fig-stdforecast}
\end{figure}  


\subsection{Fiducial forecast}

We repeat the
process of building and inverting the Fisher matrix in order to
present these forecasts as a function of $i$, as shown in
Figure~\ref{fig-stdforecast}.  The main features are:
\begin{itemize}
\item The $i$-dependence is dramatic: 
face-on targets yield much more information.  This is perhaps
counterintuitive because such a target will have a featureless
velocity field, but in our idealized model such a featureless field
carries the information that the unlensed image is exactly
circular, which is most sensitive to shear.
\item The \gx\ precision is tighter than than the \gp\ precision by a
  factor of 1.5 (at $i{=}10^\circ$) to 7 (at $i{=}50^\circ$).
  This is because \gp\ inference depends crucially on prior knowledge
  of $v_{\rm max}$ and $\mu$ while \gx\ inference depends on a more
  fundamental symmetry argument. The precision of that symmetry
  argument depends, of course, on the assumption that real galaxy
  velocity fields have negligible shearlike modes, so this assumption
  is one that should be tested in further work.
\item In this high-S/N and well-resolved scenario, both shear
  components can be inferred to a precision of 0.01 or better if the
  target is nearly face-on. At 50$^\circ$ (close to a typical value for
  randomly selected targets) \gx\ can still be inferred to this
  precision but the constraint on \gp\ is less useful.
  (\S\ref{sec-evec} will show that a linear combination of \gp\ and
  $\mu$ can still be constrained at this inclination.)
  
\end{itemize}

\subsection{Dependence on Tully-Fisher prior}\label{sec-muprior}




Tightening the Tully-Fisher (TF) prior has no effect on the \gp\
forecast because the dominant source of uncertainty for \gp\ is
uncertainty in $\mu$, at least in our fiducial setup.  This raises the
question of how loose a TF prior is tolerable. We found a
$\lesssim10\%$ relative effect on \gp\ uncertainty when the prior is
loosened from 0.04 to 0.08, and
$\approx30\%$ relative effect when further
loosened to 0.16 (i.e., 16\% scatter in rotation speed at fixed
luminosity, or almost a factor of two scatter in luminosity at fixed
speed). In summary, we see a substantial effect on \gp\ when the TF
prior becomes looser than the prior on $\mu$.  Conversely, the current
level of TFR scatter is low enough that uncertainty in $\mu$ will
remain the factor driving the \gp\ uncertainty for the foreseeable
future.

Note that neither $\mu$ nor TF priors affect the \gx\
  forecast.  In our idealized model, the limiting factor on \gx\
  precision is merely the precision and resolution of the velocity
  field measurements.  This is unlikely to be the case in nature,
  where velocity fields are not perfectly orderly. An important task
  beyond the scope of this paper is to quantify the leading sources of
  \gx\ uncertainty and systematic error due to natural variations from
  this idealized model.

\subsection{Eigenvector decomposition}\label{sec-evec}

A striking feature of our results so far is the dramatic growth of
\gp\ uncertainty with inclination, from about 1.5 times the \gx\
  uncertainty at $i=10^\circ$ to about 7 times the \gx\ uncertainty at
  $i=50^\circ$.  In this subsection we show that this is largely due
to greater mixing of \gp\ and $\mu$ as $i$ increases. To better
  illustrate what happens at high inclination, we go slightly beyond
  our standard range of $10^\circ-50^\circ$ and use a very loose $\mu$
prior of $\pm1$.

\begin{figure}
\centerline{\includegraphics[width=\columnwidth]{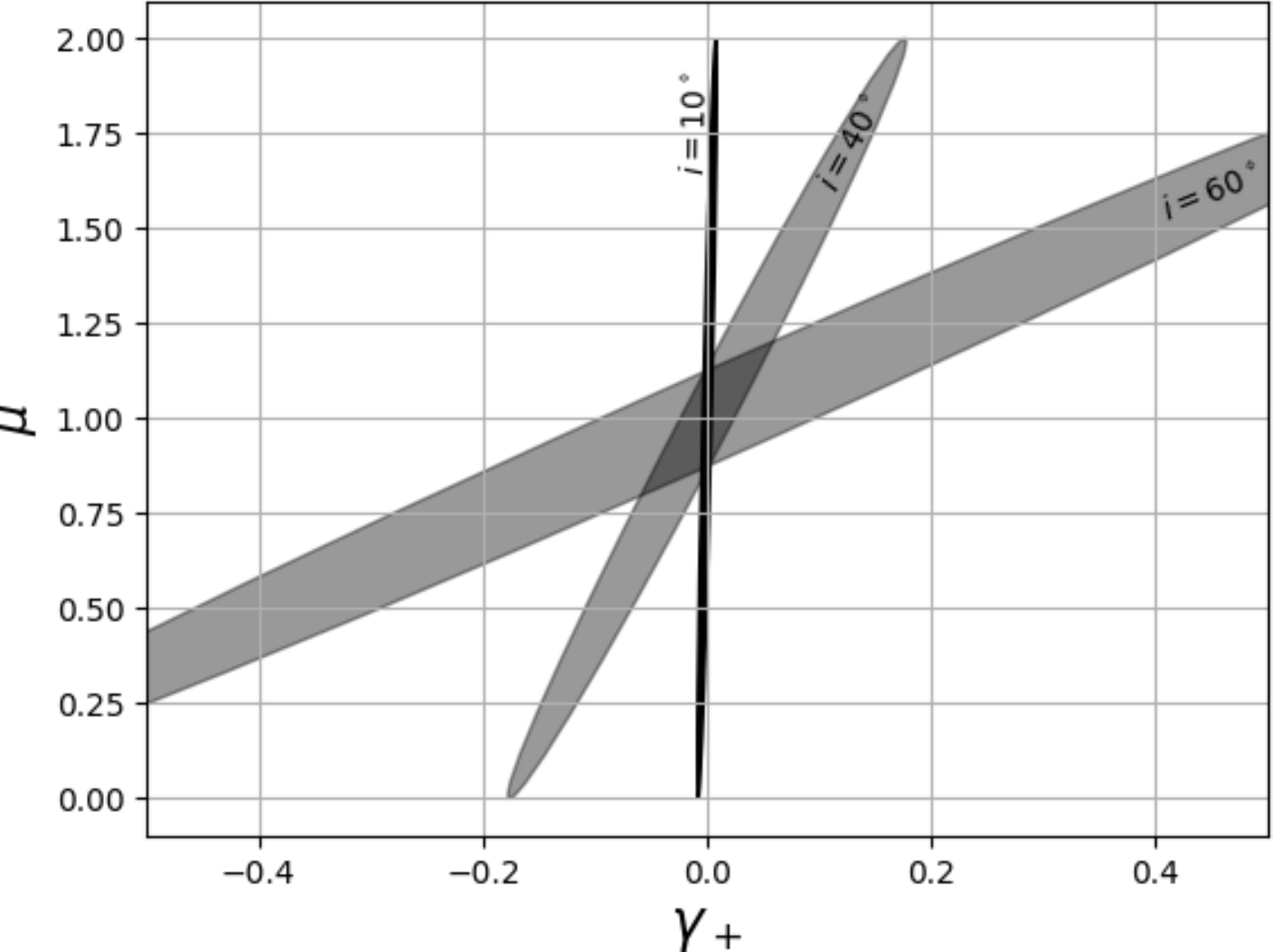}}
\caption{Constraints on \gp\ and $\mu$ as a function of inclination
  angle.}\label{fig-evec}
\end{figure}

Figure~\ref{fig-evec} illustrates the constraints in the $(\gp,\mu)$
plane at three representative inclinations.  At low inclination the
constraints on \gp\ and $\mu$ are nearly orthogonal. This makes sense
because $\mu$ should be irrelevant in the face-on case: given a
uniform velocity field, the unlensed galaxy is circular so both
components of shear can be determined precisely.  At higher
inclination, however, the constraint ellipse rotates in the
$(\gp,\mu)$ plane. With the $\mu$ uncertainty remaining $\pm1$, this
rotation greatly expands the uncertainty on \gp.

Some of the precision could be recaptured by parametrizing the lensing
in terms of eigenvectors of the $(\gp,\mu)$ submatrix of the
covariance matrix. These are represented graphically by the major and
minor axes of the ellipses in Figure~\ref{fig-evec}. Although the
minor axis does increase with $i$, it increases only about one-fifth
as much as the \gp\ uncertainty; the increase in the latter is mostly
due to the eigenvector rotation.

The eigenvector decomposition could potentially be used to improve
precision even at low inclination. Even the \gp\ width of the
$i=10^\circ$ ellipse in Figure~\ref{fig-evec} is due largely to its
$\mu$ dependence. The eigenvector decomposition defines a \gp-like
component with an uncertainty around 0.004, nearly as good as for \gx.

The practical impact of this reparametrization may depend on the
application. It may not be useful in a cosmic shear survey.  When
fitting mass profiles to lenses, however, each profile predicts both
\gp\ and $\mu$ along a given line of sight. In other words, it will
predict a point in the $(\gp,\mu)$ plane depicted in
Figure~\ref{fig-evec}. Hence, the $i=60^\circ$ ellipse may have
substantial power to discriminate between models despite the fact that
it is compatible with a range of \gp\ as well as a range of $\mu$.

Even so, it is evident that high inclinations are much less
constraining than low inclinations. Taking the inverse of the area of
this ellipse as a figure of merit, we find that the merit degrades by
a factor of six from $i=10^\circ$ to $i=40^\circ$, and by another
factor of four from there to $i=60^\circ$.

We find similar behavior when parametrizing the lensing matrix in
terms of $\kappa$ rather than $\mu$.

\subsection{Dependence on target position angle} 

In our fiducial setup the unlensed major axis position angle is
aligned with the sky coordinates ($\phi_{\rm sky}=0$), so there is no
distinction between a coordinate system fixed to the sky and one fixed
to the source galaxy.  In principle, a coordinate system fixed to the
galaxy cleanly separates the shear components into a precisely
constrained one (related to the broken symmetry of the velocity field)
and a less well constrained one (correlated with $\mu$).  In practice,
sky-based shear components must ultimately be used to interpret the
shear---to relate it to a lens, for example.  Hence we have defined
\gp\ and \gx\ based on sky coordinates. Our forecast uncertainties
have included marginalizing over the parameter $\phi_{\rm sky}$, but
our fiducial setup is still close enough to the ``pure'' galaxy basis
that \gp\ and \gx\ are clearly distinct.  In this subsection we show
that for general values of $\phi_{\rm sky}$, \gx\ is no longer an
eigenvector of the $(\gp,\gx,\mu)$ space.

\begin{figure}
\centerline{\includegraphics[width=\columnwidth]{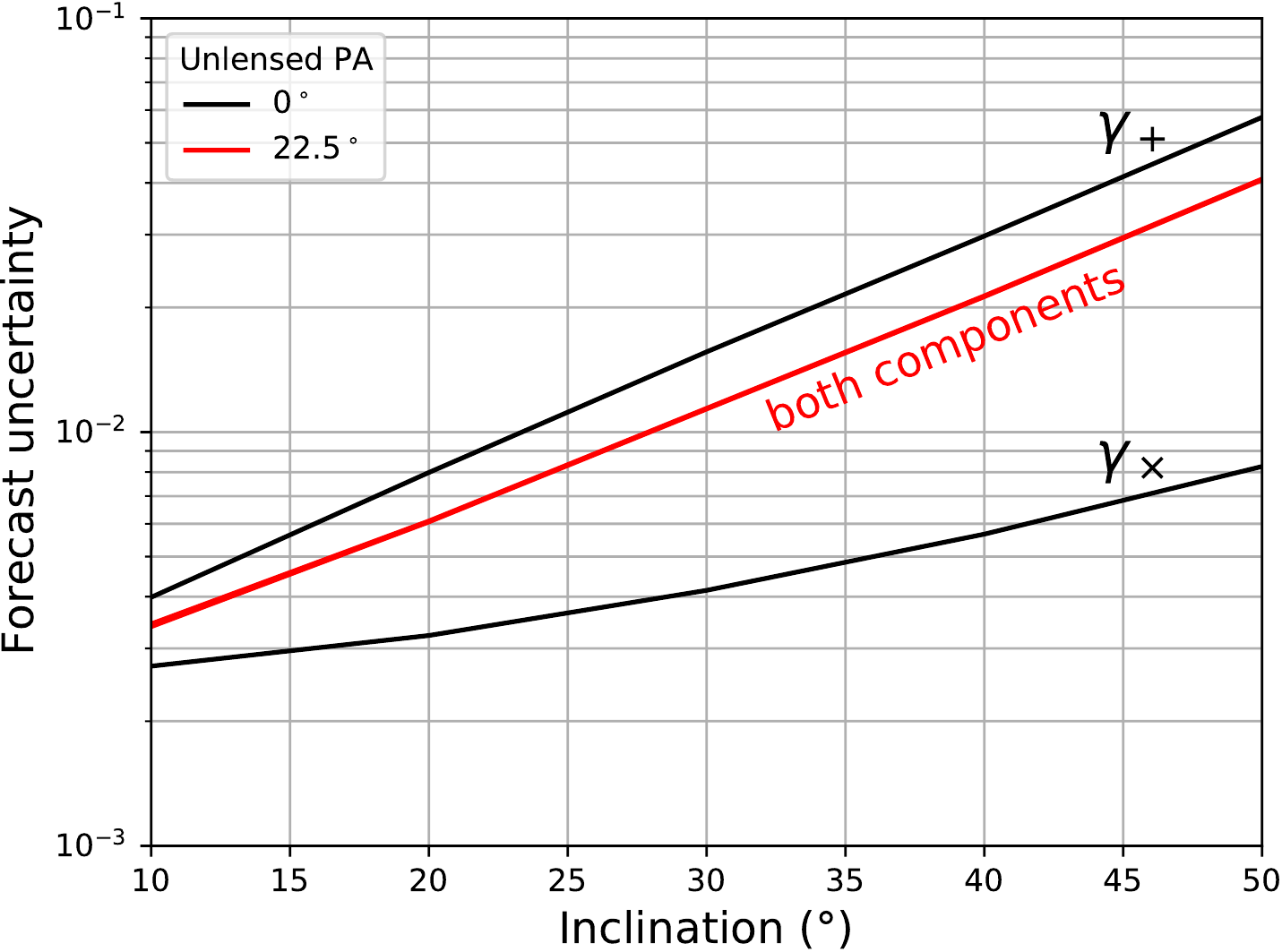}}
\caption{The effect of $\phi_{\rm sky}$ on the uncertainty of shear
  components defined on the sky. Unless $\phi_{\rm sky}$ is near a
  multiple of $45^\circ$ or the target is nearly face-on, neither
  component of shear can be measured precisely.}\label{fig-phiskydep}
\end{figure} 

Figure~\ref{fig-phiskydep} shows, along with the fiducial results, a
Fisher matrix forecast for $\phi_{\rm sky}=22.5^\circ$, where one
might expect equal precision for \gp\ and \gx. Indeed, the two red
curves representing \gp\ and \gx\ are identical and a factor of
$\sqrt{2}$ below the fiducial \gp\ forecast, indicating that the
uncertainty is maximally mixed between the two components.  If
desired, an eigenvector decomposition could be used to define two
linear combinations of \gp, \gx, and $\mu$ that are well measured
and one that is constrained only by the prior on $\mu$.  This is
just one snapshot of the $\phi_{\rm sky}$-dependent mixing: at
$\phi_{\rm sky}=45^\circ$ (not shown) \gp becomes a well-constrained
eigenvector, and there are intermediate degrees of mixing for
intermediate values of $\phi_{\rm sky}$.

A reasonable approach to forecasting precision for randomly oriented
sources would be to use the ``both components'' forecast in
Figure~\ref{fig-phiskydep}. If, for example, we are concerned with the
tangential shear of sources scattered around an axisymmetric lens, the
per-source precision will vary between the \gp\ and \gx\ curves in
Figure~\ref{fig-phiskydep}, with a mean given by the ``both
components'' curves. In this case, targets do need to be close to
face-on to reach 0.01 precision; this could be limiting, as only 13\%
of randomly oriented disks will be within 30$^\circ$ of face-on.  In
other applications it may be possible to extract more information
using the eigenvector decomposition.

We remind readers that the low uncertainty for two eigenvectors stems
from the idealized assumption that disk galaxies intrinsically have no
shearlike modes (their intrinsic major and minor axes are equal, and
their velocity fields are symmetric and locked to their intensity
fields).  To the extent that real galaxies depart from these
assumptions, the noise floor for the eigenvectors will be higher,
rendering the eigenvector decomposition (and the velocity-field method
overall) less advantageous.

\subsection{Dependence on shear} 

\begin{figure}
\centerline{\includegraphics[width=\columnwidth]{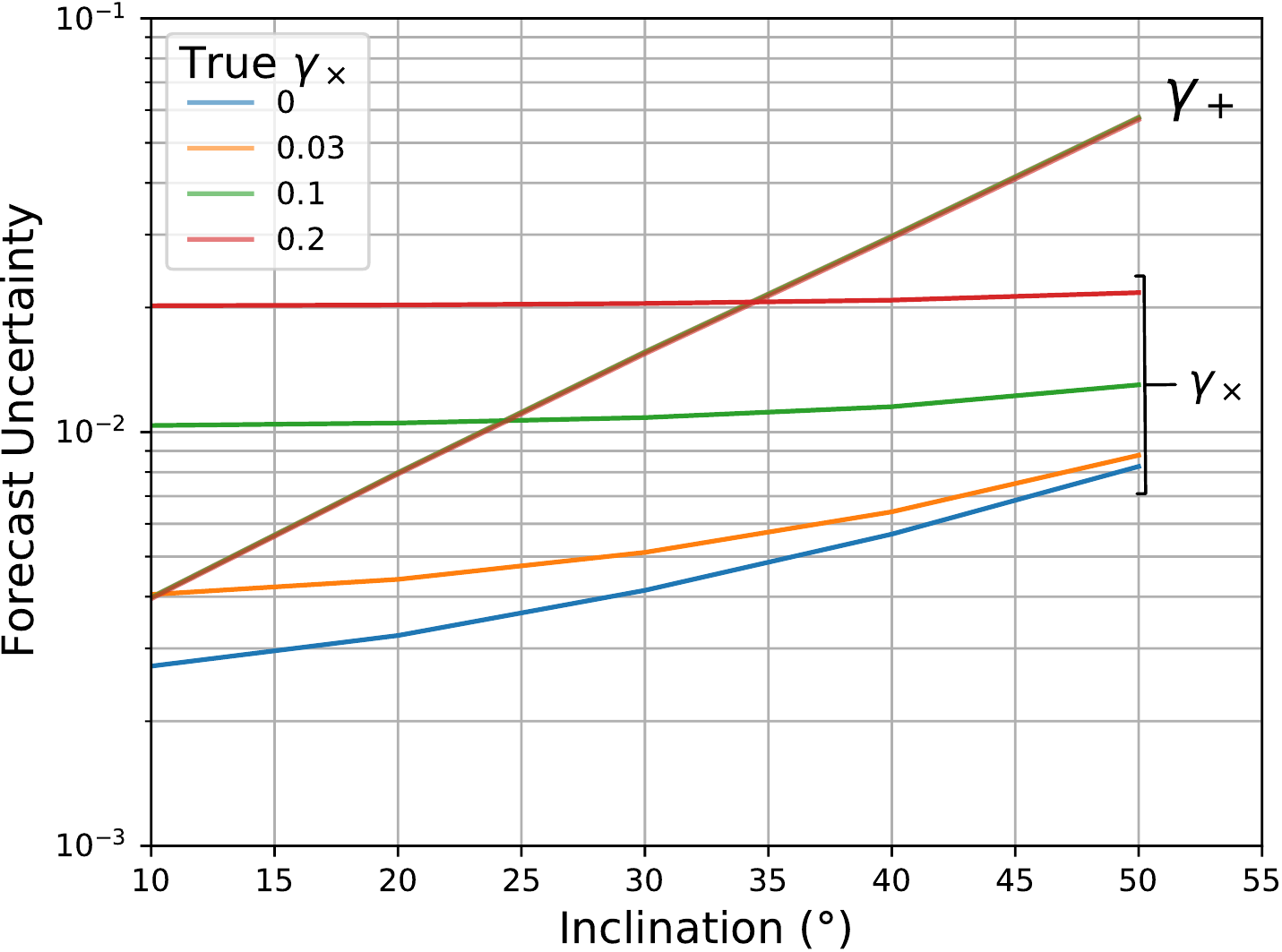}}
\caption{The uncertainty on \gx\ increases when the true \gx\ is
  substantial, due to increasing correlation with $\mu$ and related
  parameters. In the presence of substantial \gx\ the \gx\ uncertainty
  scales with both \gx\ and the size of the $\mu$ prior, which is
  $\pm0.1$ here. The \gp\ uncertainty is, in contrast, nearly
  unaffected by the true level of \gx. The true \gp\ is fixed at zero
  in this figure.}\label{fig-gammaxdep}
\end{figure}

Because the observed velocity field is not a linear function of shear,
we expect the forecast precision to depend on the shear itself.
Figure~\ref{fig-gammaxdep} presents the shear constraint forecast for
increasing levels of \gx, with \gp\ held fixed at zero.  As the true
\gx\ increases, the \gx\ precision degrades while the \gp\ precision
is nearly unaffected. The degradation occurs particularly at low
inclination where the \gx\ precision had been excellent, so the \gx\
forecast becomes more nearly independent of inclination.  In fact, as
the true \gx\ becomes substantial the \gx\ parameter becomes
correlated with $\mu$ and its family of correlated parameters. As a
result, the forecast \gx\ uncertainty scales with the $\mu$ prior as
well as the true \gx. For low true \gx, though, the \gx\ uncertainty
hits an inclination-dependent floor.  This is consistent with the
\citet{deBurghDay2015} result that $\gamma_\times\approx 0.01\pm0.01$
for two nearby galaxies with presumably negligible shear, while
suggesting that such precision is unobtainable along lines of sight
with $\gx\gtrsim0.1$.

\begin{figure}
\centerline{\includegraphics[width=\columnwidth]{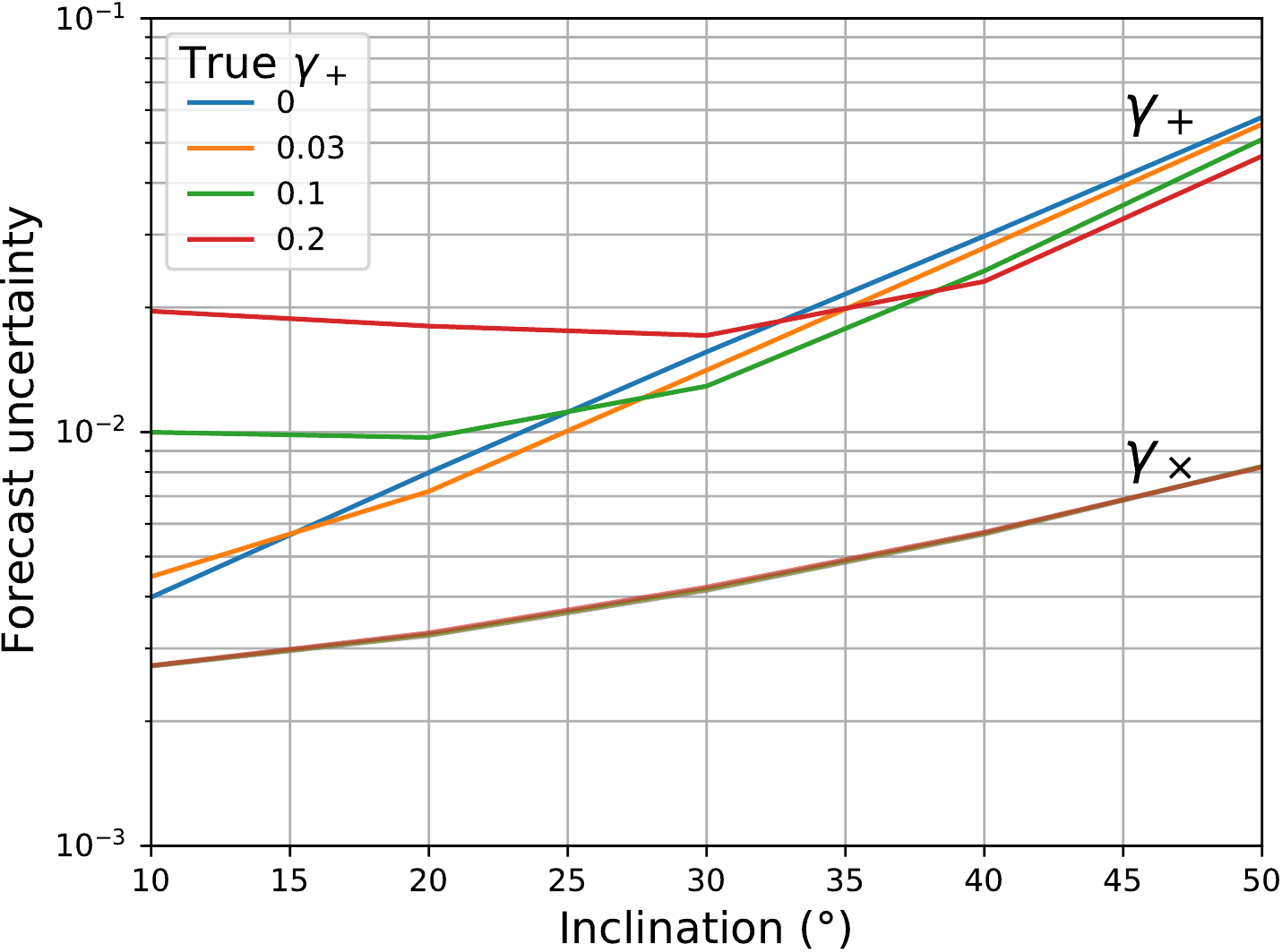}}
\caption{The \gp\ uncertainty depends on the true \gp, but much less
  dramatically than the \gx\ uncertainty depends on the true \gx\
  (Figure~\ref{fig-gammaxdep}). The \gx\ uncertainty is nearly unaffected by
  by the true level of \gp.  The true \gx\ is fixed at zero in this
  figure.}\label{fig-gammaplusdep}
\end{figure} 

Figure~\ref{fig-gammaplusdep} shows the effect of varying the true
\gp: the \gp\ precision can change in either direction by factors of a
few to several depending on the inclination, but there is no dramatic
overall trend. There is no effect on the \gx\ precision.

We also tested scenarios with mixed shear.  The precision of each
component appears to depend only on the true amount of that component,
regardless of the true amount of the other component.

These patterns can also be understood in terms of the eigenvector
decomposition. The presence of shear alters the mixing between \gp,
\gx, and $\mu$, and $\mu$ contamination is particularly noticeable in
cases where the precision had been excellent (for \gx\ generally, and
for \gp\ at low $i$). The presence of \gp\ actually {\it reduces} the
correlation with $\mu$ at higher inclinations and thus improves \gp\
inference there, but a factor of $\approx2$ reduction from a fairly
high baseline looks less dramatic on a logarithmic plot.

The eigenvector decomposition may enable useful lensing constraints in
high-shear regions. Not evident in Figures~\ref{fig-gammaxdep} and
\ref{fig-gammaplusdep} is the fact that the eigenvalues are nearly
identical regardless of shear. An eigenvector composed mostly of \gx\
but with an admixture of \gp\ and $\mu$ can still be constrained to
high precision even at $\gx=0.1$.  For fitting mass models, this could
still be highly constraining as explained in \S\ref{sec-evec}.

\subsection{Crossed slits}

\begin{figure}
\centerline{\includegraphics[width=\columnwidth]{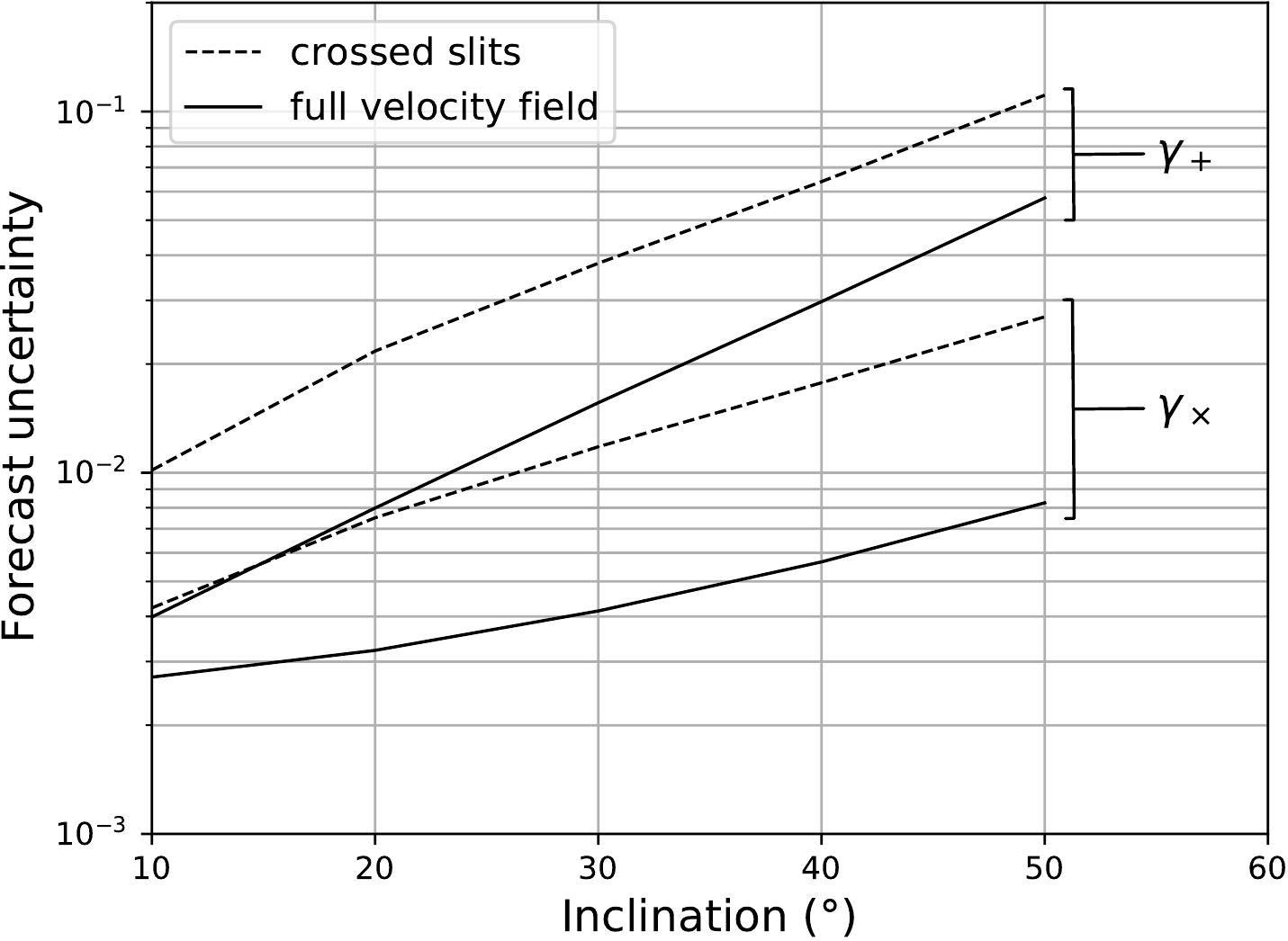}}
\caption{Shear constraints with full velocity-field observations
  (solid) versus crossed slits along the major and minor axes
  (dashed).}\label{fig-crossedslits}
\end{figure}

Obtaining full velocity-field data can be expensive, so we investigate
the suggestion of \citet{Huff2013} that slit spectra be taken across
the apparent major and minor photometric axes. (In our fiducial case
with zero shear, these are the same as the velocity-field axes.)  We
implement this by masking out most pixels in the velocity-field
partial derivative fields. In Figure~\ref{fig-crossedslits} we plot
the crossed-slit forecast as dashed curves, along with the standard
full-field forecast as solid curves. The crossed slits increase the
uncertainty by a factor of 2--3, depending only slightly on
inclination.

Although crossed slits appear to perform well for nearly face-on
targets, further work is required to assess their robustness. For
  example, consider a galaxy whose velocity field departs from our
  idealized model in a few places due to substructures, one of which
  falls in a slit. To compare the robustness of the full velocity
  field and the crossed slits across a variety of realistic galaxies,
  simulations will be required.

\subsection{Rotation Curve Model}\label{sec-URC}

We also made forecasts with the Universal Rotation Curve
\citep[URC;][]{Persic96,Salucci2007} model, which links $r_{80}$ to
the radius at which the rotation curve becomes flat; in this case
$r_0$ is still an independent parameter describing the steepness of
the rise, which can lead to an overshoot.  We found that the URC model
leads to a small improvement when nearly edge-on (ie when the rotation
curve is most apparent in observations) but otherwise yields
remarkably similar shear constraints.  We attribute this
rotation-curve insensitivity to the basic mechanisms underlying the
inference of each shear component. Inference of \gx\ relies on a
symmetry argument that should be insensitive to the specific form of
the rotation curve. Inference of \gp\ is limited by lack of knowledge
of $\mu$, a factor which is in no way ameliorated by adopting the URC
model.

\subsection{Dependence on resolution and signal-to-noise}\label{subsec-res}

Our fiducial setup uses velocity and intensity fields with 625
independent pixels (25 square) which, along the apparent major axis,
just encloses $r_{80}=12.5$ pixels.  We deliberately made our forecast
agnostic as to the target redshift and instrument details, but for
context, a typical disk scale length is about 4 kpc
\citep{2010MNRAS.406.1595F}. This yields $r_{80}\approx 12$ kpc, so
one can think of each fiducial pixel as representing about 1
kpc. Sources behind lenses are likely to have redshifts $\approx 0.4$
and up, hence their angular diameter distances set a scale of about
5--8 kpc per arcsecond (this applies to arbitrarily high redshift
sources, due to the broad maximum in angular diameter distances as a
function of redshift).  Therefore, a 1 kpc pixel will subtend 0.1--0.2
arcsec.

{\it Seeing.} To this point we have assumed that each pixel is completely
independent, but most instruments are designed with pixel sizes
smaller than the point-spread function (PSF).  Therefore we tested the
effect of blurring the intensity and velocity fields with a Gaussian
of $\sigma=2$ pixels.  This degrades the Fisher matrix forecast by
only about 10\% in relative terms. Hence, useful observations of the
lowest-redshift targets may be possible from the ground with excellent
seeing or with low-order adaptive optics, while high-redshift targets
are better pursued from space or from the ground with good adaptive
optics systems.

{\it Resolution.} We tested the effect of halving or doubling the
angular resolution, with the field size still just enclosing $r_{80}$.
For simplicity, we will describe results only for the favorable
inclination $i=10^\circ$.  Doubling (halving) the resolution
  halved (doubled) the forecast uncertainty on each shear component.
  Uncertainty decreases as pixels are added, roughly following the
  trend $n_{\rm pix}^{-1}$ where $n_{\rm pix}$ is the number of pixels
  encompassing $\pm r_{80}$ across the source major axis (and equaling
  the square root of the total number of pixels).

\begin{figure}
\centerline{\includegraphics[width=\columnwidth]{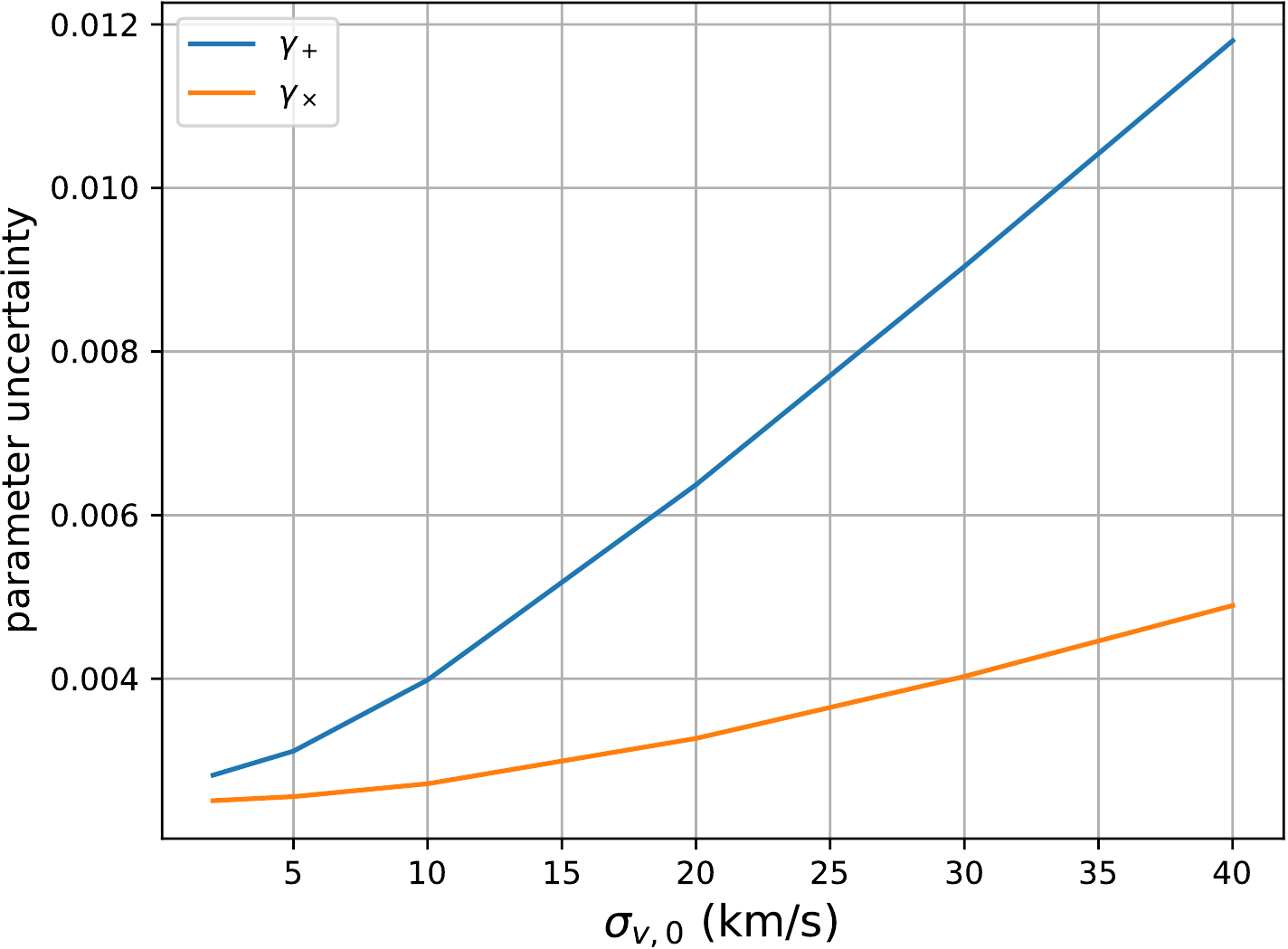}}
\caption{Shear constraints as a function of the velocity measurement
  uncertainty in the central pixel, at
  $i=10^\circ$.}\label{fig-sigmav}
\end{figure}

  {\it Velocity precision.} Figure~\ref{fig-sigmav} shows the effect
  of varying $\sigma_{v,0}$, the uncertainty in the velocity
  measurement of the central pixel, at $i=10^\circ$. Both curves can
  be well fit by a simple model in which a term depending linearly on
  $\sigma_{v,0}$ is added in quadrature to a constant noise floor;
  however, the constants depend on the component and the
  parametrization.  The \gx\ component is relatively insensitive to
  $\sigma_{v,0}$, while the \gp\ component degrades more quickly.
  Nevertheless, Figure~\ref{fig-sigmav} shows that good constraints
  can still be obtained, at least at favorable inclinations, with
  lower-precision velocity data than assumed in our fiducial model. A
  reasonable target is 10--20 km/s: efforts to go below that meet with
  diminishing returns. Sacrificing additional velocity precision
  (i.e., exposure time) to allow targeting of more galaxies may also
  be a reasonable strategy.

{\it Intensity S/N}. We reduced $I_0$, the S/N of the central
intensity pixel, from its fiducial value of 90. We found that the
precision scales nearly inversely with this S/N.

{\it Summary.} The forecast precision at $i=10^\circ$ scales roughly
as $\frac{90}{I_0}\frac{25}{n_{\rm pix}}$ where $I_0$ is the S/N of
the central intensity pixel and $n_{\rm pix}$ is the number of pixels
encompassing $\pm r_{80}$ across the source major axis. The dependence
on velocity measurement uncertainty is not so simply captured, but 10--20
km/s is a reasonable goal for observations, with little motivation to
push below that.  As a caveat, we have not investigated the
  extent, if any, to which constraints on $\mu$ degrade with data sets
  less precise than our fiducial one, hence we have not quantified the
  effect this may in turn have on shear constraints. Simulations will
  be required to address this issue.

\section{Summary and Discussion}\label{sec-discussion}

Our approach has been to assume idealized and well-measured
($\sigma_{v,0}=10$ km/s) velocity fields in order to explore the
potential of velocity field lensing. Our fiducial result is that at
the favorable inclination $i=10^\circ$ the \gx\ constraint can reach
$0.003\frac{90}{I_0}\frac{25}{n_{\rm pix}}$ where $I_0$ is the S/N of
the central intensity pixel and $n_{\rm pix}$ is the number of pixels
encompassing $\pm r_{80}$ across the source major axis. The \gp\
constraint is 1.5--7 times looser, depending on inclination.
Both constraints degrade substantially at higher inclinations.

In more detail, we find:
\begin{itemize}

\item The model is degenerate under infinitesimal displacements
    in specific directions in parameter space. However, the data
    change quadratically with step size along these directions, so the
    data can constrain all parameters.    The quadratic effect can be
    emulated in the Fisher matrix formalism by putting a $\pm0.1$
    prior on $\mu$.  
  
\item For our fiducial zero-shear scenario, constraints on \gx\ are
  precise to better than 0.01 for targets inclined by less than
  $\approx 55^\circ$---nearly half of all randomly inclined disks.
  This precision is a useful benchmark because it is roughly 20 times
  better than the per-galaxy precision for standard weak lensing, and
  also matches that found by \citet{deBurghDay2015}.  This precision,
  if true for both shear components, would make one velocity-field
  target worth roughly $20^2=400$ galaxy images, thus providing strong
  motivation to obtain the more expensive velocity-field observation.

\item This precision is more difficult to reach for \gp, the shear
  component parallel/perpendicular to the unlensed apparent major
  axis. With our default prior on $\mu$ ($\pm0.1$) only targets with
  $i<25^\circ$ reach 0.01 precision on \gp. This is a small minority
  of randomly inclined disks. Furthermore, for this select group of
  targets the assumption of face-on circularity is likely to be
  crucial, and bears further investigation.

\item For either component, constraints degrade with increasing
  $i$. For \gp\ the trend is somewhat steeper so targets with
  substantial inclination become uninteresting. The precision can be
  improved somewhat if a tighter prior on $\mu$ can be
  justified.

\item the Tully-Fisher relation is not a limiting factor.  A
  fractional velocity scatter smaller than the prior on $\mu$ is
  sufficient.


\item The notion of a well-measured \gx\ and a less well-measured \gp\
  is useful for conceptual understanding, but for general source PA
  the result is more complicated.  Of the three parameters
  $(\gp,\gx,\mu)$ two eigenvectors can be well measured and the third
  is constrained only by the prior on $\mu$. In the fiducial case,
  \gx\ is an eigenvector but the \gp-like eigenvector includes a $\mu$
  component, hence constraints on \gp\ look worse.  As $i$ increases
  that eigenvector rotates to include more $\mu$, so the pure \gp\
  constraints degrade more rapidly than the \gx\ constraints. If one
  chooses to measure the \gp-like eigenvector rather than \gp, the
  constraints degrade somewhat less rapidly with $i$.

\item In the presence of shear, the nominal \gp\ and \gx\ constraints
  degrade, but this is due to eigenvector rotation in the
  $(\gp,\gx,\mu)$ space. The eigenvalues are equally well constrained
  in the presence or absence of shear.

\item A per-pixel velocity uncertainty of 10--20 km/s is adequate,
  with smaller uncertainties yielding only marginal improvements.

\item Observing a subset of the velocity field via crossed slits may
  be a viable strategy for reducing observing expense. In the fiducial
  case ($\phi_{\rm sky}=0$) this causes a factor of 2--3
    degradation in the precision of each component.  A more realistic
  assessment of crossed slits versus full velocity fields will require
  exploration of how real disk galaxies depart from our idealized
  assumptions as well as slit placement uncertainty.

\end{itemize}

Our model is highly idealized. It assumes:
\begin{itemize}
\item The galaxy is circular when viewed face-on.
\item The velocity field is well ordered and completely described by a
  simple analytical function. The choice of rotation curve does not
  appear to matter, but the azimuthal symmetry surely matters.
\item The velocity and intensity fields share a single inclination
  angle and PA. With the arctan rotation curve, there is no other link
  between the two fields (apart from the Tully-Fisher relation). With
  the URC, there is a link via $r_{80}$ but this does not lead to
  tigher constraints because the limiting factors lie elsewhere.  
\item The disk is infinitesimally thin. The finite thickness of
  real disks will likely loosen the constraints at higher
  inclinations, because our forecast does not account for the
  increased velocity width in each pixel nor for extinction.

\item No additional structure such as bulges, bars, or warps. Bulges
  may add noise, but bars and warps seem more concerning in terms of
  biases.  Nevertheless, \cite{deBurghDay2015} did succeed in
  inferring a plausible $\gamma_\times$ ($\approx 0.01\pm0.01$) for
  radio observations of an unlensed nearby galaxy with a prominent gas
  warp, so it is possible that warps do not disturb the velocity-field
  symmetry in the same way that shear does. On the other hand, if
    the precision cited by \cite{deBurghDay2015} is due to {\it
      typical} galaxy features, future forecasts will need to account
    for this, with ${\sim}0.01$ becoming the \gx\ noise floor.
    More work is needed to address this question.
\end{itemize}

The salience of warps may hinge on the velocity-field tracer: gas or
stars. Gas is a convenient tracer for both radio and optical
spectroscopy, but is also susceptible to inflows and outflows as
well as warps.  If this leads to the velocity equivalent of shape
noise, the velocity-field method could become much less
attractive. Stellar velocity fields are more orderly, but obtaining
velocity fields from stellar absorption lines will require much more
observational effort.


These observational choices are also tied to the question of whether
the velocity and intensity fields must come from the same tracer.  In
our model the two fields are linked by a common center, inclination,
and PA.  The fact that our forecast is sensitive to the intensity
field S/N suggests that reaching the 0.01 level does require
constraints on the disk center, inclination, and PA beyond those
derivable from the velocity field itself. Therefore, misalignments
between intensity and velocity fields are a potential source of
concern.

Recent observations indicate the potential for such misalignments.
Figure~9 of \citet{ContiniMUSE2016} compares the difference between
the kinematic PA, as extracted from observations with the MUSE
integral field spectrograph at the VLT, with the morphological PA as
extracted from HST/F814W broadband images. They find one galaxy (of
27) with a large PA difference that cannot be related to poor
resolution or by being nearly face-on (where PA is less well defined):
the source of this difference is a bar. Even among the nearly face-on
cases, they attribute some of the PA differences to structures such as
spiral arms, bars, or clumps. Similarly, \citet{WisnioskiKMOS2015}
find some significant offsets between the PA of broadband light and of
the velocity field as traced by H$\alpha$ emission with the KMOS
integral field spectrograph at the VLT.  It is possible that such
offsets would be reduced (albeit at additional observational expense)
if stars were used to trace both velocity and intensity fields. Other
potential steps to mitigate this source of error could be to model
bars and spiral arms out of the intensity field, and/or to introduce a
nuisance parameter representing the intensity-velocity PA offset and
marginalize over it.

This concludes a long list of sources of uncertainty, yet to be
quantified, that could prevent this method from being of practical
use. Yet there are substantial strengths to this method as well:
\begin{itemize}
\item At favorable inclinations, tight constraints are achievable even
  with uninformative priors on $\mu$.
\item The method may work well with fitting mass models to lenses.
  Each background source will yield a constraint that may span a range
  of \gp, \gx, and $\mu$ but is a long, narrow ellipsoid in
  $(\gp,\gx,\mu)$ space.  Because a mass model predicts, for a given
  line of sight, a unique point in that space, the
  ellipsoid is likely to be highly constraining regardless of how it
  is oriented in that space.  That said, the most highly constrained
  principal axis of this ellipsoid corresponds to our fiducial \gx\
  forecast, so this argument does not allow parameter inference better
  than our fiducial forecast. Rather, inferences that cannot take
  advantage of the eigenvectors may be limited to the precisions
  presented in Figures~\ref{fig-gammaxdep} and \ref{fig-gammaplusdep}.
\item This is a method of obtaining a high-precision shear measurement
  along a {\it single} line of sight, whereas traditional weak lensing
  enables this precision only after averaging over a large area of
  sky. These are different and potentially complementary types of
  information.  The velocity field method, for example, may yield more
  information about localized substructures, which are effective
  probes of certain aspects of dark matter \citep[see, e.g.,][for an
  overview]{LSSTDMwhitepaper2019}.
\end{itemize}

\citet{Morales2006} also argued that this method avoids some of the
major systematic errors of traditional weak lensing. For example, he
argues that the PSF is no longer a first-order contributor to
systematics.  However, our assumption that the source is well-resolved
implies that the PSF would be largely irrelevant for these sources
regardless of the method. He also argues that this method is less
susceptible to contamination by intrinsic alignments.  It is indeed
robust against scenarios in which source galaxies are aligned in the
absence of lensing, because the shear is measured {\it independently}
on each target. But there are more subtle intrinsic alignment
scenarios \citep{2004PhRvD..70f3526H}. Imagine that Galaxy A sits in a
gravitational tidal field that directly affects its velocity field by
perturbing the orbits of its stars, while Galaxy B is a background
source lensed by that gravitational tidal field. To the extent that
the velocity field perturbation in Galaxy A mimics lensing modes, it
will have an ``intrinsic shear'' that is correlated with the lensing
shear on Galaxy B. In fact, this is perhaps the most important open
question here: can an external tidal field, perhaps due to a neighbor
or satellite, induce shearlike perturbations in a disk's velocity
field?  If so, marginalizing over a range of such velocity field
models could introduce significant uncertainty.  Whatever their
origin, natural sources of uncertainty will degrade \gx\ more than
\gp\ because the \gx\ forecast currently is limited only by the
precision of the velocity measurements.

A possible extension to this method is to analyze the velocity {\it
  dispersion} field as well (which requires no additional
observations).  The dispersion field is nonuniform because the disk's
radial, tangential, and vertical dispersions contribute differently to
the line-of-sight dispersion, depending on azimuth. This yields
unlensed symmetry that differs from that of the velocity field: it
is symmetric about both major and minor axes.  However, it is unlikely
that this would contribute substantially to the Fisher information,
because the azimuthal variations are small.

\acknowledgments

This work was supported in part by NSF grants 1518246 and 1911138.  We
thank Bryant Benson, Gary Bernstein, Brian Lemaux, Hunter Martin, and
Kevin Bundy for useful discussions. We also thank the anonymous
referee for significant constructive feedback.

\appendix

\section{Verifying the Fisher forecast with an exploration of the
  likelihood surface\label{appendix}}

We employ the Markov Chain Monte Carlo (MCMC) code \texttt{emcee}
\citep{emcee2013} to sample the likelihood surface.  \texttt{emcee}
implements an affine-invariant sampling algorithm, and hence performs
well even on highly degenerate systems, provided they are not strongly
multimodal \citep{2010CAMCS...5...65G}.

For each case we generate mock data and add the fiducial amount of
noise to the velocity and intensity fields.  In addition to the
Tully-Fisher prior used in the main text ($\pm0.04$ on $A_{\rm{TF}}$)
we place flat step function priors on other parameters to keep the
model physically well defined: $A_{\rm{TF}},I_0,r_0,r_{80}\ge 0$; $i$
in $[0,90^\circ)$; $\phi_{\rm sky}$ in $[0,45^\circ)$; and
$\mu>0$. Note that we do not apply the $\pm0.1$ prior on $\mu$ used
for the Fisher forecasts; the data already constrain $\mu$ to this
level due to their quadratic dependence on steps in the degeneracy
direction. We initialize one thousand walkers in a small ball around
the correct values. We run the Markov chain for $\approx1000$
autocorrelation times as a burn in, followed by an additional
$\approx1000$ autocorrelation times to record the positions of the
walkers.

\begin{figure}
\includegraphics[width=\textwidth]{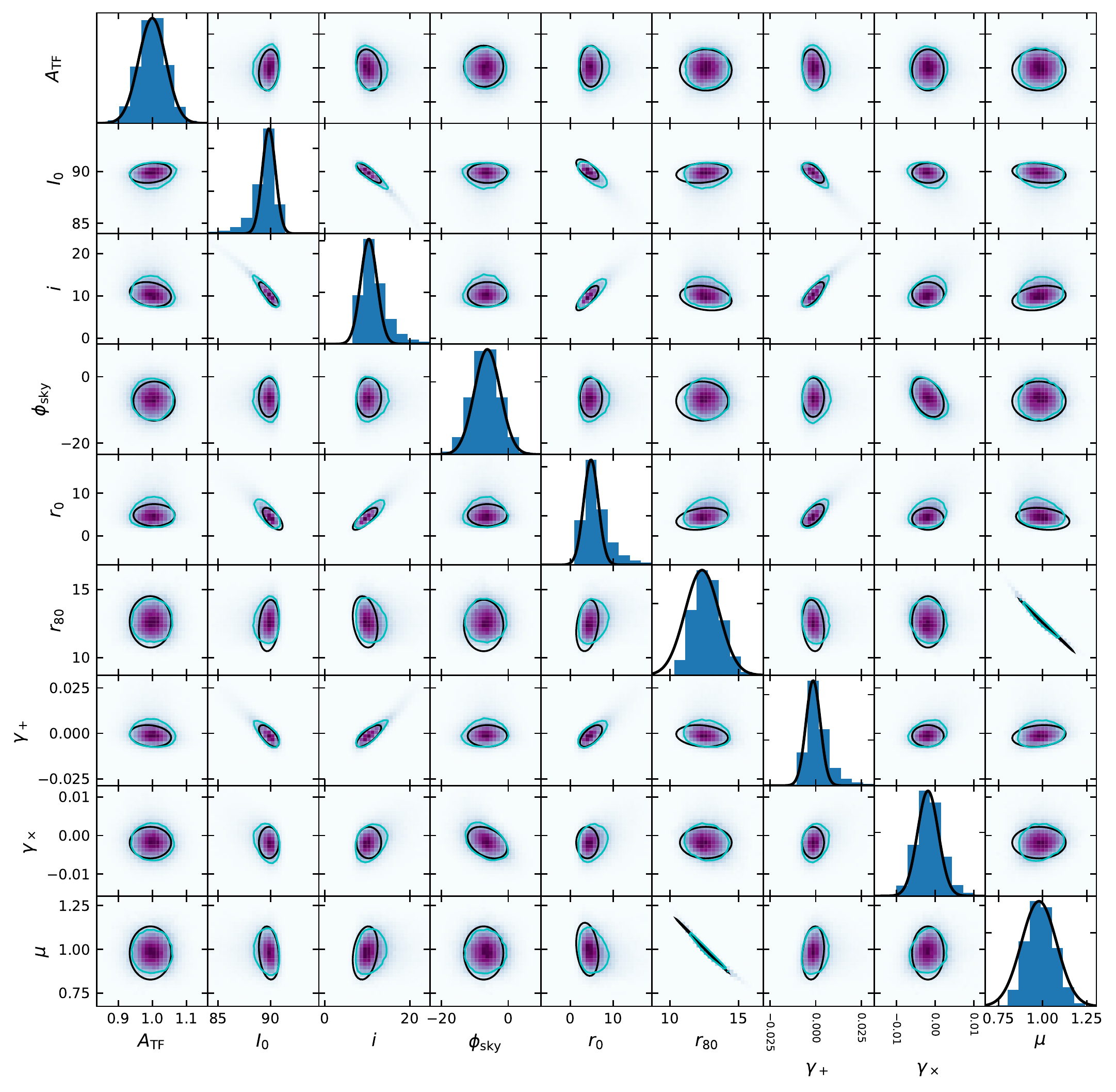}
\caption{Corner plot for the fiducial case at $i=10^\circ$, showing
  agreement between MCMC confidence intervals and the Fisher forecast.
  The colorscale represents the density of MCMC samples, the cyan
  contour represents the MCMC 68\% confidence region, and the black
  ellipse represents the Fisher forecast for the 68\% confidence
  region. The forecast ellipses are centered on the densest MCMC
  pixel.}\label{fig-MCMCi10}
\end{figure}

\begin{figure}
\includegraphics[width=\textwidth]{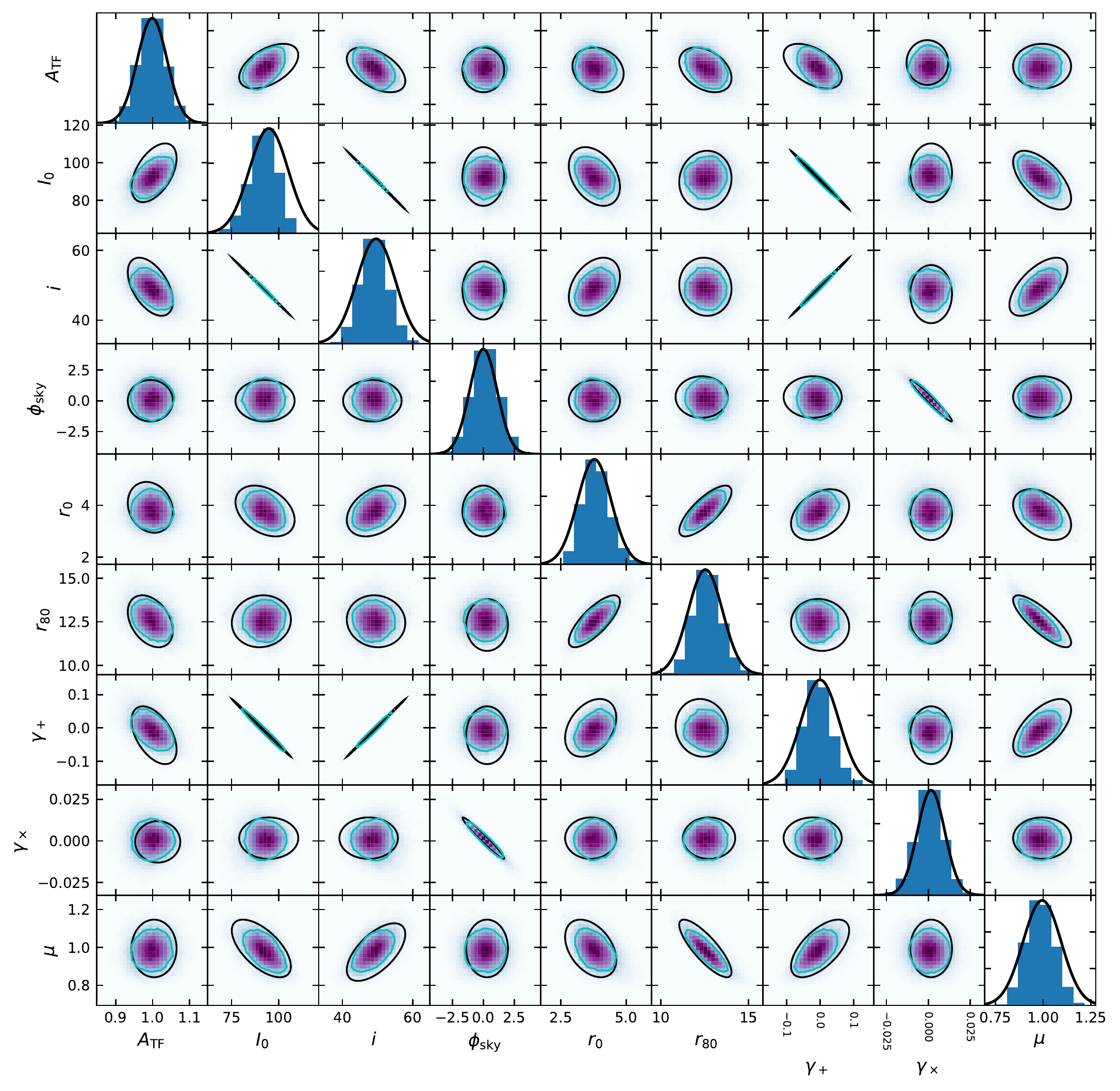}
\caption{As for Figure~\ref{fig-MCMCi10}, but at $i=50^\circ$.}\label{fig-MCMCi50}
\end{figure}

We first examine our fiducial case at the extremes of
  $i=10^\circ$ (Figure~\ref{fig-MCMCi10} and $i=50^\circ$
  (Figure~\ref{fig-MCMCi50}.  In each figure, the colorscale
  represents the density of MCMC samples, the cyan contour represents
  the MCMC 68\% confidence region, and the black ellipse represents
  the Fisher forecast for the 68\% confidence region. The three least
  interesting parameters ($x_0$, $y_0$, and $v_0$) are omitted for
  clarity.  A good match is evident throughout all panels at
  $i=10^\circ$. At $i=50^\circ$ there is a hint that the forecast is
  becoming more pessimistic than the MCMC samples. This effect is more
  noticeable at higher inclinations.  We attribute this to increasing
  numerical errors as one goes to higher inclinations: the condition
  number at $i=50^\circ$ ($60^\circ$) is $7\times10^7$
  ($3\times10^8$).  Hence, for the fiducial setup we provide forecasts
  only for $i\le 50^\circ$, and more generally we provide forecasts
  only where the condition number is ${<}10^8$.

\begin{figure}
\includegraphics[width=\textwidth]{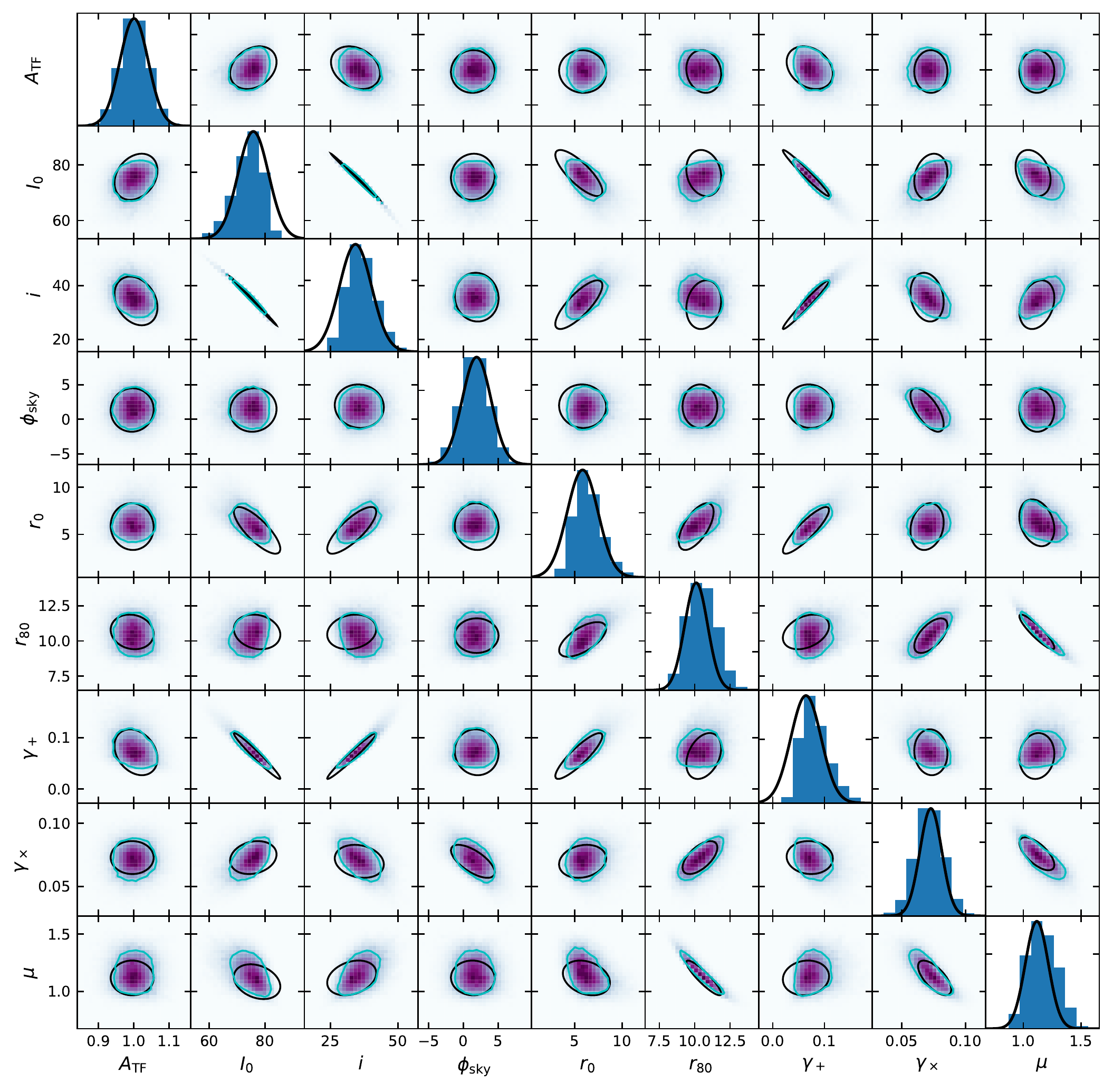}
\caption{As for Figure~\ref{fig-MCMCi10}, but for the
  far-from-fiducial setup described in the text. Note that the
  forecast contours are centered on the densest MCMC pixel, rather
  than the MCMC centroid.}\label{fig-MCMCcase1}
\end{figure}

Finally, we present a case far from our fiducial scenario: with
$\phi_{\rm sky}=10^\circ$ and $\gp=\gx=0.07$, these three parameters
are highly mixed. We have also changed $r_0$ ($r_{80}$) to 6 (10) kpc,
$I_0$ to 75, and $\mu$ to 1.2, with an inclination angle of
$35^\circ$.  Figure~\ref{fig-MCMCcase1} shows that the forecast still
accurately predicts the MCMC precision.

\bibliography{ms}

\begin{thebibliography}{}
\expandafter\ifx\csname natexlab\endcsname\relax\def\natexlab#1{#1}\fi

\bibitem[{{Bartelmann} \& {Maturi}(2017)}]{Bartelmann17}
{Bartelmann}, M., \& {Maturi}, M. 2017, Scholarpedia, 12, 32440

\bibitem[{{Blain}(2002)}]{Blain2002}
{Blain}, A.~W. 2002, \apjl, 570, L51

\bibitem[{{Blakeslee}(2001)}]{Blakeslee2001}
{Blakeslee}, J.~P. 2001, arXiv Astrophysics e-prints, astro-ph/0108253

\bibitem[{{Contini} {et~al.}(2016){Contini}, {Epinat}, {Bouch{\'e}},
  {Brinchmann}, {Boogaard}, {Ventou}, {Bacon}, {Richard}, {Weilbacher},
  {Wisotzki}, {Krajnovi{\'c}}, {Vielfaure}, {Emsellem}, {Finley}, {Inami},
  {Schaye}, {Swinbank}, {Gu{\'e}rou}, {Martinsson}, {Michel-Dansac},
  {Schroetter}, {Shirazi}, \& {Soucail}}]{ContiniMUSE2016}
{Contini}, T., {Epinat}, B., {Bouch{\'e}}, N., {et~al.} 2016, \aap, 591, A49

\bibitem[{{de Burgh-Day} {et~al.}(2015){de Burgh-Day}, {Taylor}, {Webster}, \&
  {Hopkins}}]{deBurghDay2015}
{de Burgh-Day}, C.~O., {Taylor}, E.~N., {Webster}, R.~L., \& {Hopkins}, A.~M.
  2015, \mnras, 451, 2161

\bibitem[{{Drlica-Wagner} {et~al.}(2019){Drlica-Wagner}, {Mao}, {Adhikari},
  {Armstrong}, {Banerjee}, {Banik}, {Bechtol}, {Bird}, {Boddy}, {Bonaca},
  {Bovy}, {Buckley}, {Bulbul}, {Chang}, {Chapline}, {Cohen-Tanugi}, {Cuoco},
  {Cyr-Racine}, {Dawson}, {D{\'\i}az Rivero}, {Dvorkin}, {Erkal}, {Fassnacht},
  {Garc{\'\i}a-Bellido}, {Giannotti}, {Gluscevic}, {Golovich}, {Hendel},
  {Hezaveh}, {Horiuchi}, {Jee}, {Kaplinghat}, {Keeton}, {Koposov}, {Lam}, {Li},
  {Lu}, {Mandelbaum}, {McDermott}, {McNanna}, {Medford}, {Meyer}, {Marc},
  {Murgia}, {Nadler}, {Necib}, {Nuss}, {Pace}, {Peter}, {Polin},
  {Prescod-Weinstein}, {Read}, {Rosenfeld}, {Shipp}, {Simon}, {Slatyer},
  {Straniero}, {Strigari}, {Tollerud}, {Tyson}, {Wang}, {Wechsler}, {Wittman},
  {Yu}, {Zaharijas}, {Ali-Ha{\"\i}moud}, {Annis}, {Birrer}, {Biswas}, {Blazek},
  {Brooks}, {Buckley-Geer}, {Caputo}, {Charles}, {Digel}, {Dodelson},
  {Flaugher}, {Frieman}, {Gawiser}, {Hearin}, {Hlo{\v{z}}ek}, {Jain},
  {Jeltema}, {Koushiappas}, {Lisanti}, {LoVerde}, {Mishra-Sharma}, {Newman},
  {Nord}, {Nourbakhsh}, {Ritz}, {Robertson}, {S{\'a}nchez-Conde}, {Slosar},
  {Tait}, {Verma}, {Vilalta}, {Walter}, {Yanny}, \&
  {Zentner}}]{LSSTDMwhitepaper2019}
{Drlica-Wagner}, A., {Mao}, Y.-Y., {Adhikari}, S., {et~al.} 2019, arXiv
  e-prints, arXiv:1902.01055

\bibitem[{{Fathi} {et~al.}(2010){Fathi}, {Allen}, {Boch}, {Hatziminaoglou}, \&
  {Peletier}}]{2010MNRAS.406.1595F}
{Fathi}, K., {Allen}, M., {Boch}, T., {Hatziminaoglou}, E., \& {Peletier},
  R.~F. 2010, \mnras, 406, 1595

\bibitem[{{Foreman-Mackey} {et~al.}(2013){Foreman-Mackey}, {Hogg}, {Lang}, \&
  {Goodman}}]{emcee2013}
{Foreman-Mackey}, D., {Hogg}, D.~W., {Lang}, D., \& {Goodman}, J. 2013, \pasp,
  125, 306

\bibitem[{{Garcia-Fernandez} {et~al.}(2016){Garcia-Fernandez}, {S{\'a}nchez},
  {Sevilla-Noarbe}, {Suchyta}, {Huff}, {Gaztanaga}, {Aleksi{\'c}}, {Ponce},
  {Castander}, {Hoyle}, {Abbott}, {Abdalla}, {Allam}, {Annis},
  {Benoit-L{\'e}vy}, {Bernstein}, {Bertin}, {Brooks}, {Buckley-Geer}, {Burke},
  {Carnero Rosell}, {Carrasco Kind}, {Carretero}, {Crocce}, {Cunha},
  {D'Andrea}, {da Costa}, {DePoy}, {Desai}, {Diehl}, {Eifler}, {Evrard},
  {Fernandez}, {Flaugher}, {Fosalba}, {Frieman}, {Garc{\'{\i}}a-Bellido},
  {Gerdes}, {Giannantonio}, {Gruen}, {Gruendl}, {Gschwend}, {Gutierrez},
  {James}, {Jarvis}, {Kirk}, {Krause}, {Kuehn}, {Kuropatkin}, {Lahav}, {Lima},
  {MacCrann}, {Maia}, {March}, {Marshall}, {Melchior}, {Miquel}, {Mohr},
  {Plazas}, {Romer}, {Roodman}, {Rykoff}, {Scarpine}, {Schubnell}, {Smith},
  {Soares-Santos}, {Sobreira}, {Tarle}, {Thomas}, {Walker}, \&
  {Wester}}]{DESSVmag}
{Garcia-Fernandez}, M., {S{\'a}nchez}, E., {Sevilla-Noarbe}, I., {et~al.} 2016,
  ArXiv e-prints, arXiv:1611.10326

\bibitem[{{Goodman} \& {Weare}(2010)}]{2010CAMCS...5...65G}
{Goodman}, J., \& {Weare}, J. 2010, Communications in Applied Mathematics and
  Computational Science, 5, 65

\bibitem[{{Heavens}(2016)}]{Heavens2016}
{Heavens}, A. 2016, Entropy, 18, 236

\bibitem[{{Hirata} \& {Seljak}(2004)}]{2004PhRvD..70f3526H}
{Hirata}, C.~M., \& {Seljak}, U. 2004, \prd, 70, 063526

\bibitem[{{Huff} \& {Graves}(2014)}]{HuffMagnification2014}
{Huff}, E.~M., \& {Graves}, G.~J. 2014, \apjl, 780, L16

\bibitem[{{Huff} {et~al.}(2013){Huff}, {Krause}, {Eifler}, {George}, \&
  {Schlegel}}]{Huff2013}
{Huff}, E.~M., {Krause}, E., {Eifler}, T., {George}, M.~R., \& {Schlegel}, D.
  2013, ArXiv e-prints, arXiv:1311.1489

\bibitem[{{Miller} {et~al.}(2011){Miller}, {Bundy}, {Sullivan}, {Ellis}, \&
  {Treu}}]{MillerTFR2011}
{Miller}, S.~H., {Bundy}, K., {Sullivan}, M., {Ellis}, R.~S., \& {Treu}, T.
  2011, \apj, 741, 115

\bibitem[{{Morales}(2006)}]{Morales2006}
{Morales}, M.~F. 2006, \apjl, 650, L21

\bibitem[{{Morrison} {et~al.}(2012){Morrison}, {Scranton}, {M{\'e}nard},
  {Schmidt}, {Tyson}, {Ryan}, {Choi}, \& {Wittman}}]{Morrison12}
{Morrison}, C.~B., {Scranton}, R., {M{\'e}nard}, B., {et~al.} 2012, \mnras,
  426, 2489

\bibitem[{{Persic} {et~al.}(1996){Persic}, {Salucci}, \& {Stel}}]{Persic96}
{Persic}, M., {Salucci}, P., \& {Stel}, F. 1996, \mnras, 281, 27

\bibitem[{{Press} {et~al.}(1992){Press}, {Teukolsky}, {Vetterling}, \&
  {Flannery}}]{NumRec}
{Press}, W.~H., {Teukolsky}, S.~A., {Vetterling}, W.~T., \& {Flannery}, B.~P.
  1992, {Numerical recipes in C. The art of scientific computing}

\bibitem[{{Rizzo} {et~al.}(2018){Rizzo}, {Vegetti}, {Fraternali}, \& {Di
  Teodoro}}]{Rizzo2018}
{Rizzo}, F., {Vegetti}, S., {Fraternali}, F., \& {Di Teodoro}, E. 2018, \mnras,
  481, 5606

\bibitem[{{Salucci} {et~al.}(2007){Salucci}, {Lapi}, {Tonini}, {Gentile},
  {Yegorova}, \& {Klein}}]{Salucci2007}
{Salucci}, P., {Lapi}, A., {Tonini}, C., {et~al.} 2007, \mnras, 378, 41

\bibitem[{{Tully} \& {Fisher}(1977)}]{TullyFisher}
{Tully}, R.~B., \& {Fisher}, J.~R. 1977, \aap, 54, 661

\bibitem[{{Vallisneri}(2008)}]{2008PhRvD..77d2001V}
{Vallisneri}, M. 2008, \prd, 77, 042001

\bibitem[{{Wisnioski} {et~al.}(2015){Wisnioski}, {F{\"o}rster Schreiber},
  {Wuyts}, {Wuyts}, {Bandara}, {Wilman}, {Genzel}, {Bender}, {Davies},
  {Fossati}, {Lang}, {Mendel}, {Beifiori}, {Brammer}, {Chan}, {Fabricius},
  {Fudamoto}, {Kulkarni}, {Kurk}, {Lutz}, {Nelson}, {Momcheva}, {Rosario},
  {Saglia}, {Seitz}, {Tacconi}, \& {van Dokkum}}]{WisnioskiKMOS2015}
{Wisnioski}, E., {F{\"o}rster Schreiber}, N.~M., {Wuyts}, S., {et~al.} 2015,
  \apj, 799, 209

\end{thebibliography}

\end{document}